\documentclass[a4paper,11pt]{article}

\usepackage{jheppub}
\usepackage{tikz}
\usetikzlibrary{shapes}
\usetikzlibrary{plotmarks}
\usepackage{tikz-3dplot}
\usepackage{mathtools}

\title{Contour Integral for the Partition Function of $\mathcal{N}=2$ Topologically Twisted on $\mathbb{CP}^2$ and Physical Fluxes}
\author[1,2]{Lorenzo Ruggeri}
\affiliation[1]{Dipartimento di Matematica ``Giuseppe Peano'', Università di Torino, Via Carlo Alberto 10, 10123 Torino, Italy}
\affiliation[2]{INFN, Sezione di Torino, Via Pietro Giuria 1, 10125 Torino, Italy}

\emailAdd{lorenzo.ruggeri@unito.it}

\abstract{We compute the contour integral for the partition function of an $\mathcal{N}=2$ $SU(2)$ topologically twisted theory on $\mathbb{CP}^2$, dimensionally reducing from an $\mathcal{N}=1$ theory on $S^5$. Earlier works presented the partition function as a sum over three equivariant fluxes, one for each toric divisor of $\mathbb{CP}^2$. Our result depends only on a single physical flux, assigned to the non-trivial two-cycle of the manifold. The reduced summation over fluxes is compensated by a contour of integration, arising from a different solution of the BPS equations, which captures more poles in each topological sector. As our observable involves a position-dependent Yang-Mills coupling, we compute new equivariant invariants of $\mathbb{CP}^2$, which reduce to Donaldson invariants in the non-equivariant limit. Stability conditions of gauge bundles over $\mathbb{CP}^2$ appear intrinsically via the dimensional reduction.}

\begin{document}

\maketitle

\section{Introduction}
Supersymmetric localization \cite{Witten:1988ze,Nekrasov:2002qd,Pestun:2007rz} is a powerful technique that allows to reduce the integral over all field configurations to a finite-dimensional integral over a set of zero-modes. The Abelian effective theory governing the zero-modes can be computed exactly via localization, and it receives both perturbative and non-perturbative contributions.  Since the integrand displays a rich pole structure, determining the integration contour is often an arduous task.  This work focuses on $\mathcal{N}=2$ topologically twisted pure $SU(2)$ gauge theories on $\mathbb{CP}^2$, for which the integral is over zero-modes for the complex scalar in the vector multiplet. The effective theory consists of topological sectors labelled by instantons and fluxes. In particular, we show how the choice of contour and the restriction over flux sectors play a crucial role in computing the integral over zero-modes.

The zero-modes one integrates over are determined by solving the BPS equations. A prototypical example of this interplay is provided by the localization on $S^2$ of an $\mathcal{N}=(2,2)$ vector multiplet coupled to matter with gauge group $SU(2)$. In \cite{Benini:2012ui,Doroud:2012xw}, a BPS solution is found in which one component of the complex scalar is covariantly constant, while the second component is proportional to the gauge flux. Hence, to take into account zero-modes for the covariantly constant scalar, the integral is defined over the real line. To compute the integral one closes the contour at infinity and computes the residue at each pole of the integrand. Instead, for the A-twisted theory on $S^2$ \cite{Closset:2015rna} both components of the scalar are covariantly constant. Thus, the integral is over the complex plane from which one removes the neighbourhoods around certain points in the space of zero-modes where massless modes of the chiral fields may develop. By showing that the Abelian effective theory is a total derivative, the integral is reduced to the boundaries around the singular regions and it picks up poles at those points. In this case, the result of the integration is expressed in terms of a geometrical operation called Jeffrey-Kirwan (JK) residue \cite{Jeffrey:1993cun,Benini:2013nda,Hori:2014tda}\footnote{A similar interplay between BPS solution and integration over zero-modes appears in 3d for the $\mathcal{N}=2$ superconformal \cite{Kim:2009wb,Imamura:2011su,Kapustin:2011jm} and topologically twisted index \cite{Benini:2015noa}.}. Another important example is given by $\mathcal{N}=2$ $SU(2)$ theories on $S^3$. Here, the contour can be expressed as an integral over zero-modes either only for the real scalar \cite{Kapustin:2009kz,Hama:2011ea,Jafferis:2010un} or also including the gauge connection along the Hopf fiber \cite{Closset:2017zgf}. The JK residue prescription in the latter case is shown to reproduce the so-called $\sigma$-contour in the former case \cite{Closset:2017zgf}.

Returning to the topologically twisted theory on $\mathbb{CP}^2$, earlier results in the literature \cite{Bershtein:2015xfa,Bershtein:2016mxz,Bonelli:2020xps} found the space of zero-modes at a given topological sector, the \emph{Coulomb branch}, to be two-dimensional for a pure $SU(2)$ gauge theory. This arises since, once evaluated on the BPS locus, both components of the complex scalar in the $\mathcal{N}=2$ vector multiplet $(a,\bar{a})$ are covariantly constant. To guarantee maximal symmetry breaking, the point $a=0$ is removed from the integral over the Coulomb branch. Similarly to the JK residue above, the effective Abelian theory is shown to be a total derivative and only the residue at the origin is picked up. By summing over all flux sectors, the evaluation of the residue sum passes several non-trivial consistency checks as it reproduces wall-crossing formulas \cite{Gottsche:2006tn} and, in the non-equivariant limit, Donaldson invariants \cite{ellingsrud1995wallcrossing}. However, it has been recently observed \cite{Kim:2025fpz} that this procedure does not agree with the expected one from a JK residue computation.

In this work\footnote{While our focus is on gauge group $SU(2)$, our procedure can be straightforwardly extended to $SO(3)$ gauge theories and thus to the full $U(2)$ gauge theories as in \cite{Bershtein:2015xfa,Bershtein:2016mxz,Bonelli:2020xps}.}, we find a seemingly different solution to the BPS equations where, as for the non-twisted theory on $S^2$ \cite{Benini:2012ui,Doroud:2012xw}, one component of the scalar is fixed by the flux sector\footnote{As we will motivate momentarily, this solution arises naturally when dimensionally reducing from an $\mathcal{N}=1$ theory on $S^5$.}. Hence, we find that the integral is one dimensional and, by closing the integral at infinity, we pick up an infinite amount of poles of the integrand for each flux sector. Despite this difference, when we sum over all fluxes we reproduce\footnote{Up to a position-dependent coupling constant to be discussed shortly.} the results in \cite{Bershtein:2015xfa,Bershtein:2016mxz,Bonelli:2020xps}. To understand how such different integration contours can give the same result, one has to consider how the sum over fluxes is treated in the two setups.

If the spacetime manifold has non-trivial two-cycles, one expects gauge field configurations carrying flux to contribute to the partition function of supersymmetric gauge theories\footnote{\label{fn:1}Such partition functions are obtained by applying supersymmetric localisation. In order to solve the BPS equations, one typically chooses a gauge such that the two real scalars take values in the Cartan subalgebra of the gauge group. It was shown in \cite{Blau:1994rk} that, even for the trivial gauge field configuration, this ``diagonalisation'' of the scalars can be obstructed. In order for this obstruction to be lifted, one has to include torus gauge bundles into the BPS locus, and the corresponding gauge fields have flux on two-cycles.}. There are two different proposals to account for such contributions in the case where the spacetime is a 4d toric manifold: one is in terms of equivariant flux assigned to each toric divisor \cite{Nekrasov:2003vi,Bawane:2014uka,Bershtein:2015xfa,Bershtein:2016mxz,Bonelli:2020xps}, while the other one is in terms of ``physical'' flux assigned to each non-trivial two-cycle in the manifold \cite{Lundin:2021zeb,Lundin:2023tzw,Ruggeri:2025kmk}. Clearly, then, the sum over equivariant fluxes in the partition function is redundant in comparison with the sum over physical fluxes. In our work we show that the different sums over topological sectors precisely compensates the difference in the amount of poles picked up by the integration contour in the two cases. Hence, our main result shows that localizing an $\mathcal{N}=2$ topologically twisted theory on $\mathbb{CP}^2$ can be done in two inequivalent ways leading to the same final expression\footnote{Uplifting the two theories on $\mathbb{CP}^2\times S^1$ and exploiting the fibering operators \cite{Closset:2018ghr,Closset:2022vjj}, these two localizations formulas can be applied to an $\mathcal{N}=1$ theory on $S^5$. Under this operation, the equivariant and non-equivariant two-cycles in $\mathbb{CP}^2\times S^1$ are mapped, respectively, to contractible equivariant and non-equivariant three-cycles in $S^5$. In this setup the two localization formulas could possibly be related to the two inequivalent description of giant gravitons in \cite{Imamura:2021ytr} and \cite{Gaiotto:2021xce} as D3 branes wrapping, respectively, equivariant and non-equivariant three-cycles on $S^5$.}.

A crucial aspect of our result is that it is obtained dimensionally reducing an $\mathcal{N}=1$ theory on a squashed $S^5$ \cite{Kallen:2012va,Imamura:2012efi,Kim:2012qf,Lockhart:2012vp} along its Hopf fiber to the $\mathbb{CP}^2$ base \cite{Lundin:2021zeb,Lundin:2023tzw}. Instead of performing dimensional reduction shrinking the radius of the $S^1$-fiber, we first act with a $\mathbb{Z}_h$-quotient. At finite $h$, the resulting manifold is a lens space and, as it is not simply connected, the partition function is a sum over topological sectors labelled by the winding number $\mathfrak{m}$ of the non-trivial flat connections. At large $h$, the fiber shrinks to a point and the non-trivial flat connections give rise to gauge configurations with flux on the two-cycle of\footnote{See \cite{Benini:2012ui} for an earlier computation where the partition function of an $\mathcal{N}=(2,2)$ theory on $S^2$ has been shown to arise taking the large $h$ limit of the partition function on a lens space.} $\mathbb{CP}^2$. Hence, our 4d partition function only depends on the physical flux $\mathfrak{m}$, rather than a triplet of equivariant fluxes, one for each toric divisor of $\mathbb{CP}^2$. 

The contour of integration similarly arises via dimensional reduction. On the lens space, the component of the gauge field along the fiber direction is set to be proportional to the winding number $\mathfrak{m}$ of the flat connection. Upon dimensional reduction, the component of the 4d scalar arising from the gauge field in 5d is proportional to the flux sector $\mathfrak{m}$. Therefore, we only integrate over a one-dimensional space, corresponding to the covariantly constant scalar descending from the 5d scalar. Importantly, the 5d theory requires the scalar to be Wick rotated in order to have a positive kinetic term and stronger localization locus \cite{Qiu:2016dyj}. Hence, we integrate over the imaginary line. As this contour crosses al poles of the integrand, we deform it and close it at (real) infinity. 

An interesting aspect of the theory obtained via dimensional reduction is that the 4d Yang-Mills coupling $\tilde{g}_\text{4d}$ is position dependent. This arises because the length of the Hopf fiber, which enters the relation between $g_\text{5d}$ and $\tilde{g}_\text{4d}$, depends on the point of the base \cite{Festuccia:2016gul}. Hence, the observable for which we compute the partition function in 4d is not the one appearing in \cite{Bershtein:2015xfa,Bershtein:2016mxz,Bonelli:2020xps}. Then, the evaluation of the partition function provides new equivariant invariants of $\mathbb{CP}^2$. In the non-equivariant limit, which consists in starting from a round $S^5$, the 4d coupling is actually constant. We show that our observable reproduces Donaldson invariants \cite{ellingsrud1995wallcrossing} in this limit. Note, however, that both one-loop determinants and instanton contribution of the Abelian effective theory can be precisely mapped to those in \cite{Bershtein:2015xfa,Bershtein:2016mxz,Bonelli:2020xps}. Thus, by switching observable, it is possible to use our integration contour and sum over fluxes to reproduce the equivariant Donaldson invariants computed there.

Our approach also shows that stability conditions of gauge bundles on $\mathbb{CP}^2$ arise naturally for a generic $SU(N)$ gauge group. As we mentioned, allowed fluxes arise from non-trivial flat connection with winding number $\mathfrak{m}=\diag(\mathfrak{m}_1,\mathfrak{m}_2-\mathfrak{m}_1,\dots,-\mathfrak{m}_{N-1})$. At a finite $\mathbb{Z}_h$-quotient the winding numbers are restricted to be $0\leq\mathfrak{m}_i<h$. Hence, for large $h$ we find that only positive fluxes $\mathfrak{m}_i$ enter in the partition function. Moreover, in each topological sector at a finite $\mathbb{Z}_h$-quotient, the charges under rotation along the fiber of the modes need to satisfy a projection condition $t=\alpha(\mathfrak{m})\mmod h\geq 0$ \cite{Alday:2012au}. At large $h$ this imposes $\alpha(\mathfrak{m})\geq 0$. Together, these two conditions provide the stability conditions for gauge bundles on $\mathbb{CP}^2$. 

The outline is as follows. In \autoref{sec.2} we present the dimensional reduction from a 5d $\mathcal{N}=1$ theory on a toric Sasakian manifold to its 4d quasi-toric base. We then specify in \autoref{sec.3} to the $S^1$-quotient of $S^5$, namely $\mathbb{CP}^2$. We also study the analytic structure of the integrand and compare with the approach in \cite{Bershtein:2015xfa,Bershtein:2016mxz,Bonelli:2020xps}. Our main result appears in \autoref{sec.4}, where, exploiting properties of the residue sum, we compute our equivariant topological invariants on $\mathbb{CP}^2$. We also show that they reduce to Donaldson invariants in the non-equivariant limit. Finally, in \autoref{sec.5}, we summarize our main results and present a list of interesting directions for future research.

\section{Review of Partition Function on Quasi-Toric Four-Manifolds} \label{sec.2}
In this section we review how the partition function of certain $\mathcal{N}=2$ pure gauge theories on closed, connected and simply-connected four-manifolds are computed. Specifically, we consider theories obtained by a procedure devised in \cite{Festuccia:2018rew,Festuccia:2019akm} which generalises Witten's topological twist \cite{Witten:1988ze} and Pestun's theory \cite{Pestun:2007rz} on the four-sphere in the following way: given 4d $\mathcal{N}=2$ SYM on a four-manifold $M$ and a Killing vector field $v$ generating an isometry on $M$, one can choose, w.l.o.g., an atlas where each chart contains exactly one fixed point of $v$ (since $M$ is compact). Upon a choice of assigning ``$+$'' or ``$-$'' to each chart, the output of this procedure is an equivariant cohomological field theory, whose BPS gauge field is anti-self-dual (ASD) on the ``$+$'' charts and self-dual (SD) on the ``$-$'' charts. In other words, the theory localises to instantons on the ``$+$'' charts and anti-instantons on the ``$-$'' charts. Equivariant Donaldson-Witten theory is obtained as the special case where one assigns ``$+$'' to all charts, resulting in the topological twist\footnote{We stress that equivariance is essential for this procedure and the resulting cohomological theory seizes to exist in the non-equivariant limit---the exception being the topological twist.}.

If we consider such cohomological theories on spacetimes that contain non-trivial two-cycles, then the main difficulty is to obtain flux contributions to the partition function (which we expect for reasons discussed in footnote \ref{fn:1}). A method to obtain such contributions was devised in \cite{Lundin:2021zeb,Lundin:2023tzw}. The idea is to start with an $\mathcal{N}=1$ pure gauge theory in 5d and obtain the cohomological theory in 4d via dimensional reduction. For this purpose we require the corresponding 5d spacetime $M$ to be a connected, simply-connected toric Sasakian manifold and a (non-trivial) $S^1$-fibration over the 4d spacetime $B$:
\begin{equation}\label{eq.fibration}
    \begin{tikzcd}
    S^1\ar[r,hook] & M\ar[d,"\pi"]\\ & B
    \end{tikzcd}
\end{equation}
Dimensional reduction proceeds in two steps: first, one considers the 5d theory $\mathcal{T}_{M/\mathbb{Z}_h}$ on a finite quotient of $M$ along the fibre-direction. The resulting spacetime is no longer simply-connected but has finite fundamental group. This introduces non-trivial flat gauge connections to the BPS locus of $\mathcal{T}_{M/\mathbb{Z}_h}$ that have to be accounted for in the partition function $\mathcal{Z}[\mathcal{T}_{M/\mathbb{Z}_h}]$. The reduction to 4d is performed by taking $h$ to be large and the corresponding partition function is obtained as $\lim_{h\to\infty}\mathcal{Z}[\mathcal{T}_{M/\mathbb{Z}_h}]$. In this limit, the flat connections in the locus of $\mathcal{T}_{M/\mathbb{Z}_h}$ precisely account for flux configurations in the locus of the 4d theory \cite{Lundin:2023tzw}. Let us now state the main steps of this procedure in some detail.

\subsection{Five-Dimensional Gauge Theory}\label{sub.5d}

The study of $\mathcal{N}=1$ pure gauge theory on toric Sasakian five-manifolds $M$ was introduced in \cite{Kallen:2012cs}; see \cite{Qiu:2016dyj} for a review. In particular, $M$ is equipped with a contact structure $\mathcal{S}$ whose contact form\footnote{We assume that both $M$ and  $\mathcal{S}$ are orientable, such that $\mathcal{S}=\ker\kappa$ globally. Note that $\kappa$ does not determine $\mathcal{S}$ uniquely.} we denote by $\kappa$. Moreover, we can define the unique vector field $\R$ such that
\begin{equation}
    \iota_\R\kappa=1,\qquad \iota_\R\dd\kappa=0,
\end{equation}
called the Reeb vector field. In order for the contact structure to be compatible with the metric $g$ of $M$, we also require $L_{\R}g=0$ and $g(u,v)=\frac{1}{2}\dd\kappa(u,J(v))$, where $u,v\in\ker\kappa$ and $J$ an almost complex structure on $\ker\kappa$. By virtue of being toric, $M$ is in one-to-one correspondence with a good, convex, rational polyhedral cone \cite{Lerman:2001zua}, which is precisely the moment map cone---denoted by $\mathcal{C}$ in the following---of its metric cone $C(M)$: $\mathcal{C}$ plays a crucial role for the partition function of $\mathcal{T}_M$. The Sasakian property implies that there exists a Kähler structure transverse to $\R$, which facilitates the one-loop computation.

The fields of a twisted vector multiplet consist of a gauge field $A$, a real bosonic scalar $\sigma$ and a fermionic one-form $\psi$ transforming in the adjoint representation under the gauge group $G$. Off-shell closure of the supersymmetry algebra also requires an auxiliary fermionic two-form $\chi$ and a bosonic two-form $H$. They both satisfy a horizontal self-duality condition with respect to the projector $P^+=\frac{1}{2}(1+\iota_\R\star)$. Supersymmetry acts on the fields as follows:
\begin{equation}\label{eq.5dSUSY}\setlength{\tabcolsep}{1.5em}
    \begin{tabular}{ll}
    $\delta A = i\Psi$, & $\delta \Psi = -\iota_{\R} F+ \dd_A\sigma$,\\[1ex]
    $\delta\chi = H$, & $\delta H = -iL_{\R}^A\chi-[\sigma,\chi]$,\\[1ex]
    $\delta\sigma = -i\iota_{\R}\Psi$, &
    \end{tabular}
    \end{equation} 
with $\delta^2=-\ii L_{\R}+T_\Phi$, where $T_\Phi$ denotes a gauge transformation by $\Phi=\sigma+\iota_{\R}A$. 

The supersymmetric action for the 5d $\mathcal{N}=1$ vector multiplet consists of an observable and a $\delta$-exact part:
\begin{equation}\label{eq.action5d}
    S=\int_M\frac{1}{g_\text{5d}^2}\tr\big(-CS_{3,2}(A+\sigma\kappa)-\ii(\kappa\wedge \dd\kappa\wedge\Psi\wedge\Psi)+\delta \mathcal{V}\big),
\end{equation}
where
\begin{align}
    &CS_{3,2}(A)=\kappa\wedge F\wedge F,\\ 
    &\mathcal{V}=\Psi\wedge\star(-\iota_{\R} F-\dd_A\sigma)-\frac{1}{2}\chi\wedge\star H+2\chi\wedge\star F+\sigma\kappa\wedge \dd\kappa\wedge\chi
\end{align}
and $g^2_{5d}$ is the Yang-Mills coupling in 5d. The partition function of this theory can be computed via localisation, by adding an additional term
\begin{equation}
    t\int\delta\left(\Psi\wedge\star(-\iota_{\R} F-\dd_A\sigma)-\frac{1}{2}\chi\wedge\star H+2\chi\wedge\star F\right)
\end{equation}
to the action with $t$ a positive constant (sent to infinity in the process of localising). After Wick-rotating $\sigma, H$ (in order to obtain positive kinetic terms), the localisation locus is given by \cite{Qiu:2016dyj}
\begin{equation}\label{eq.5dBPS}
    P^+F=0,\qquad\dd_A\sigma=0.
\end{equation}
On the simply-connected spacetime $M$, non-trivial solutions for the first equation are given by contact instantons \cite{Baraglia_2016}; there are no non-trivial flat connections. In the perturbative sector, the gauge connection is trivial and $\sigma$ is a constant, Lie algebra-valued scalar, denoted by $a\in\ii\mathfrak{g}$.

\paragraph{Partition Function.}
Once gauge-fixing has been performed, the partition function receives three contributions: a classical term obtained by evaluating $\exp(-S)$ on the localisation locus, a perturbative term obtained by computing the superdeterminant of $-\ii L_\R+\ii T_a$, and a non-perturbative term accounting for contact instantons. The full expression was conjectured
\cite{Qiu:2016dyj} to be\footnote{In this expression a sum over topological sectors labelled by gauge fluxes, one for each non-trivial two-cycle in $H_2(M,\mathbb{Z})$, is missing. This issues has been addressed in \cite{Ruggeri:2025kmk} for $Y^{p,q}$ and $L^{a,b,c}$. Note that $S^5$, the main focus of this paper, does not have non-trivial two-cycles.}
\begin{equation}\label{eq.Z5d}
    \mathcal{Z}[\mathcal{T}_M]=\int_{\ii\mathfrak{t}}\dd a\; \mathrm{e}^{-\frac{8\pi^3r^3}{g_\text{YM}^2}\rho\tr a^2}{\det}'_\text{adj}\,S^{\mathcal{C}}_3(\ii a|\R) \prod_{\ell=1}^n Z^\text{inst}_{\mathbb{C}^2\times S^1}(\ii a|\beta_\ell,\epsilon^\ell_1,\epsilon^\ell_2).
\end{equation}
Here, $\rho=\text{Vol}_M/\text{Vol}_{S^5}$, the primed determinant over the adjoint representation excludes zero-eigenvalues (corresponding to zero-modes of $\sigma$ along the root $\alpha$) and $S^\mathcal{C}_3$ denotes the generalised triple sine function
\begin{equation}\label{eq.1loop.M}
    S^\mathcal{C}_3(x|\R)=\prod_{\vec{n}\in\mathcal{C}\cap\mathbb{Z}^3}(\vec{n}\cdot\vec{\R}+x)\prod_{\vec{n}\in\mathring{\mathcal{C}}\cap\mathbb{Z}^3}(\vec{n}\cdot\vec{\R}-x).
\end{equation}
Here, $\vec{n}=(n_1,n_2,n_3)$ represents the charges of the modes under the $T^3$-action and $\mathring{\mathcal{C}}$ denotes the interior of $\mathcal{C}$.
Finally, $Z^\text{inst}_{\mathbb{C}^2\times S^1}$ denotes Nekrasov's instanton partition function on $\mathbb{C}^2\times S^1$ with parameters $\beta_\ell,\epsilon_1^\ell,\epsilon_2^\ell$. 
Explicitly, let $\vec{v}_\ell$ be the inward-pointing normals of the moment polytope of $M$ and $\vec\X$ such that\footnote{At this point, the choice of $\vec{\X}$ is arbitrary. Its significance is related to the theory in 4d and will be discussed shortly.} $\vec\X\cdot(\vec v_\ell\times\vec v_{\ell+1})=\pm 1$, then  
\begin{equation}\label{eq.equivariance.loc} 
\beta_\ell=\frac{\vec\X\cdot(\vec v_\ell\times\vec v_{\ell+1})}{\vec\R\cdot(\vec v_\ell\times\vec v_{\ell+1})},\qquad\epsilon_1^\ell=\frac{\vec\X\cdot(\vec\R\times \vec v_{\ell+1})}{\vec\X\cdot(\vec v_\ell\times \vec v_{\ell+1})},\qquad\epsilon_2^\ell=\frac{\vec\X\cdot(\vec v_\ell\times\vec\R)}{\vec\X\cdot(\vec v_\ell\times\vec v_{\ell+1})}.
\end{equation} 
These are, respectively, the radius of the Reeb orbit and the (local) equivariance parameters for the $T^2$-action on $\mathbb{C}^2$ at the $\ell$th vertex of the moment polytope of $M$.

\subsection{Dimensional Reduction}
In order to obtain from \eqref{eq.Z5d} a partition function on the four-dimensional base $B$ of the principal $S^1$-bundle \eqref{eq.fibration}, we perform a dimensional reduction along the $S^1$-fibre. For toric Sasakian $M$, vector fields $\X$ generating a free $S^1$-action on $M$ can be found as follows: write $\X=\sum_{i=1}^3\X_ie_i$, where $\{e_i\}$ is a basis of vector fields generating the $T^3$-action on $M$. Then $\X$ must solve the equations
\begin{equation}\label{eq.freeS1}
    \vec\X\cdot(\vec v_\ell\times\vec v_{\ell+1})=\pm 1
\end{equation}
for all $\ell$. Note in particular that such free directions $\X$ are generally different from the Reeb vector field $\R$, which is free only when $M$ is regular. 

Equipped with the vector field $\X$ generating the free $S^1$-action in \eqref{eq.fibration}, let us also define the one-form $b:=g(\X/\|\X\|^2,\;\cdot\;)$, such that $\iota_\X b=1$ ($g$ is the metric on $M$). Then we can rewrite the fields of $\mathcal{T}_M$ as
\begin{equation}\label{eq.fields.5d4d}\setlength{\tabcolsep}{1.5em}
    \begin{tabular}{ll}
    $A=\pi^\ast A_4+\varphi_4 b$, & $\psi=\pi^\ast\psi_4+\eta b$,\\[1ex]
    $\chi=\pi^\ast\chi_4+b\,\iota_\X\chi$, & $H=\pi^\ast H_4+b\,\iota_\X H$,\\[1ex]
    $\sigma=-\ii\phi_4-\iota_\R b\;\varphi_4$
    \end{tabular}
\end{equation}
and restricting to gauge transformations which are independent of the fibre direction. The 4d field content is precisely the one of the 4d theories on $B$ studied in \cite{Festuccia:2018rew,Festuccia:2019akm}, where the assignment of ``$+$'' or ``$-$'' to the chart containing the $i$th fixed point is determined by the choice of sign in \eqref{eq.freeS1}. Note in particular that $\chi,H\in\Omega^{2+}_H(M)$ implies $\chi_4,H_4\in P^+_\omega\Omega^2(B)$, where $P^+_\omega$ is the projector\footnote{Briefly, \cite{Festuccia:2018rew} defines a smooth function $\omega$ on $B$ which assumes values $\pm\pi$ at each fixed point, depending on the assignment of ``$+$'' or ``$-$'' to the corresponding patch. Then the projector defined as $P^+_\omega=\frac{1}{1+\cos^2\omega}(1+\cos\omega\star-\sin^2\omega\frac{\kappa\wedge\iota_v}{\|v\|^2})$ projects to forms in $\Omega^{2+}$ on patches where $\omega=\pi$ and to forms in $\Omega^{2-}$ on ones where $\omega=-\pi$.} in \cite{Festuccia:2018rew,Festuccia:2019akm} and $\cos\omega=g(\R,\X/\|\X\|)$. The 4d Killing vector field utilised for the 4d theories of \cite{Festuccia:2018rew,Festuccia:2019akm} is obtained as $v:=\pi_\ast\R$.
        
By plugging \eqref{eq.fields.5d4d} into the 5d SUSY algebra \eqref{eq.5dSUSY}, one can see that the 4d fields satisfy precisely the SUSY algebra of \cite{Festuccia:2018rew,Festuccia:2019akm}. Consequently, BPS solutions of our 5d theory reduce to BPS solutions of the 4d theory there\footnote{We will comment momentarily about the comparison with the BPS solution in \cite{Bershtein:2015xfa}.}. It can also be shown that the deformation complex used to compute the one-loop determinant in \cite{Festuccia:2018rew,Festuccia:2019akm} follows directly from dimensional reduction of the complex obtained for the 5d theory. Finally, the contact instanton equation in \eqref{eq.5dBPS} reduces to (a deformation of) either instanton or anti-instanton equations for each chart in 4d, depending on the sign of $\iota_\R b$ \cite{Festuccia:2016gul,Festuccia:2019akm}. We conclude that the one-loop determinant and instanton terms of the 5d $\mathcal{N}=1$ pure gauge theory precisely determine the one-loop determinant and instanton terms of the theories studied in \cite{Festuccia:2018rew,Festuccia:2019akm} resulting after dimensional reduction.

\paragraph{Finite Quotients of $M$.}
The way we perform dimensional reduction from $M$ to $B$ is, as mentioned above, by considering the 5d theory on finite quotients of $M$ along the $S^1$-fibre, $M/\mathbb{Z}_h$, and subsequently send $h\to\infty$. This procedure was introduced in the context of the 4d theories studied here in \cite{Lundin:2021zeb,Lundin:2023tzw}. Locally, $\mathcal{T}_M$ and $\mathcal{T}_{M/\mathbb{Z}_h}$ are the same; however, due to $\pi_1(M/\mathbb{Z}_h)\simeq\mathbb{Z}_h$, the localisation locus of $\mathcal{T}_{M/\mathbb{Z}_h}$ includes non-trivial flat connections. For gauge group $G=SU(N)$, these are (locally and up to gauge transformations) of the form $A=\mathfrak{m}\,\dd\alpha$, where $\alpha$ is the angle of the $S^1$-fibre and $\mathfrak{m}=\diag(\mathfrak{m}_1,\mathfrak{m}_2-\mathfrak{m}_1,\dots,-\mathfrak{m}_{N-1})$ any element of the Cartan subalgebra $\mathfrak{t}$ such that $\mathfrak{m}_i\in\mathbb{Z}_{\ge0}$, $0\le\mathfrak{m}_i< h$ for all $i$ and $\mathfrak{m}_i\le\mathfrak{m}_{i+1}$ for $i=1,\dots,N-2$. Solutions for the scalar $\sigma$ are still given by constants $a\in\ii\mathfrak{g}$, however, such that $[\mathfrak{m},a]=0$ (i.e., generically $a\in\ii\mathfrak{t}$). In order to obtain the full partition function $\mathcal{Z}[\mathcal{T}_{M/\mathbb{Z}_h}]$ we have to sum over all the topological sectors, labelled by $\mathfrak{m}$.

What is the interpretation of $\mathfrak{m}$ in the limit where $h\to\infty$? Using the relations in \eqref{eq.fields.5d4d}, flat connections in 5d imply 
\begin{equation}
    0=F=\pi^\ast F_4+\dd_{\pi^\ast A_4}\varphi\wedge b+\varphi\,\dd b.
\end{equation}
Furthermore, when going around the $S^1$-fibre in the background of a flat connection, fields will pick up a holonomy of the form
\begin{equation}
    \exp\left(\ii\int A\right)=\mathrm{e}^{2\pi\ii\mathfrak{m}/h},
\end{equation}
from which we deduce
\begin{equation}\label{eq.varphibps}
    \varphi_4=\mathfrak{m}.
\end{equation}
Consequently, we find\footnote{Note that $\dd b$ is basic with respect to $\X$, i.e. $L_\X\dd b=0$ and $\iota_\X\dd b=0$.} 
\begin{equation}
    F_4=-\mathfrak{m}\,\dd b.
\end{equation}
Note that $[\dd b]$ is non-trivial in $H^2(B)$ and, by the nature of the fibration \eqref{eq.fibration}, there exists a non-trivial two-cycle $[c]\in H_2(B)$ such that
\begin{equation}
    \frac{1}{2\pi}\int_cF_4=\mathfrak{m}.
\end{equation}
Thus, in the limit $h\to\infty$, the flat connections of the 5d theory give rise to field strength saddles carrying flux in the 4d theory.

Let us now state the results of the reduction for the three different parts of $\mathcal{Z}[\mathcal{T}_{B}]$.

\paragraph{Classical Part.}
The classical part is obtained by evaluating \eqref{eq.action5d} on the saddles of $\mathcal{T}_{M/\mathbb{Z}_h}$ and then sending $h\to\infty$. A peculiarity of the 5d reduction is that the 4d Yang-Mills coupling at the $\ell$th fixed point, $\tilde{g}^2_{4d,\ell}$, is position-dependent whenever $\X$ is not proportional to $\R$. The two couplings are related as\footnote{For ease of notation, we set $r=1$ for the remainder of this work.}
\begin{equation}\label{eq.coupling}
    \tilde{g}^2_{\text{4d}}=\frac{g^2_\text{5d}}{2\pi\|\X\|}h,
\end{equation}
the dependence on $h$ appears as the length of the fibers goes as $1/h$; see \cite{Festuccia:2016gul,Lundin:2023tzw} for details. We denote the 4d coupling, which is actually constant, as $g^2_\text{4d}=\frac{g^2_\text{5d}}{2\pi}h$.

Taking the large $h$ limit, we obtain\footnote{In the limit $h\rightarrow\infty$ we keep the product $g^2_\text{5d}\cdot h$ fixed.}
\begin{equation}\begin{split}\label{eq.classical.4d}
    Z_B^\text{cl}&=\exp\left(-\lim_{h\to\infty} \int_{M/\mathbb{Z}_h} CS_{3,2} (A_\text{flat}+a\kappa)\right)\\
    &=\exp\left(\sum_\ell \frac{(2\pi)^2}{\tilde{g}^2_{\text{4d},\ell}}
    \frac{\tr (\ii a+\iota_\R b\,\mathfrak{m})^2}{\epsilon_1^\ell\epsilon_2^\ell}\right)\\
    &=\exp\left(-\frac{(2\pi)^2\rho}{g^2_\text{4d}} \tr(a)^2\right),
\end{split}\end{equation}
for each topological sector $\mathfrak{m}$. The sum on the right-hand side is over fixed points of the Killing vector field $v$ and the local equivariance parameters have been defined in \eqref{eq.equivariance.loc}.

Note that the classical term \eqref{eq.classical.4d} is different from the one used in \cite{Festuccia:2018rew,Festuccia:2019akm} (see (58) in \cite{Festuccia:2018rew}) precisely because of the position-dependence of $\tilde{g}^2_\text{4d}$. In particular, all terms involving $\mathfrak{m}$ in \eqref{eq.classical.4d} cancel among each other, since the non-trivial flat connections on $M/\mathbb{Z}_h$ do not contribute to the classical action \eqref{eq.action5d}.

\paragraph{One-Loop Part.}
It was shown in \cite{Lundin:2021zeb,Lundin:2023tzw} that the one-loop determinant of $\mathcal{T}_{M/\mathbb{Z}_h}$, for a given topological sector $\mathfrak{m}$ and root $\alpha$ in the root set $\Delta$ of $G$, can be obtained from \eqref{eq.1loop.M} simply by restricting the products to slices $\mathcal{C}_t=\{\vec u\in\mathcal{C}|\langle\vec u,\vec\X\rangle=t\}\subset\mathcal{C}$ of the moment map cone\footnote{Intuitively, this can be understood as follows: the one-loop determinant \eqref{eq.1loop.M} essentially counts holomorphic functions on the metric cone $C(M)$, weighted by their charge under $L_\R$. Naively, on the quotient space one would expect that only holomorphic functions whose charge under $L_\X$ is a multiple of $h$ survive. However, since the quotient space admits flat connections, labelled by $\mathfrak{m}$ and valued in a subset of $\mathfrak{t}$, we can allow for holomorphic sections of charge $\alpha(\mathfrak{m})\mmod h$. For more details see \cite{Lundin:2023tzw}; for the index computation see \cite{Qiu:2016dyj}.}. Here, $t$ labels the charge of the modes for a rotation along the fiber, and it is related to the topological sector by the \emph{projection condition}
\begin{equation}\label{eq.projection}
    t=\alpha(\mathfrak{m})\mmod h.
\end{equation}
For the topological twist, $\mathcal{C}_{t<0}=\emptyset$, as $\X$ is contained in the moment map cone.

For large $h$, the projection condition simplifies to
\begin{equation}\label{eq.projection.B}
    t=\alpha(\mathfrak{m})
\end{equation}
and it determines which of the integral points inside the moment map cone $\mathcal{C}$ contribute to the one-loop determinant at flux sector $\mathfrak{m}$. Consequently, in a fixed topological sector we obtain the one-loop contribution
\begin{equation}\label{eq.1loop.B}
    Z_B^\text{1-loop}=\lim_{h\to\infty} \prod_{\alpha\in\Delta}\prod_{\substack{t\in\mathbb{Z}\\t=\alpha(\mathfrak{m})\mmod h}}S^{\mathcal{C}_t}_3(\ii a|\R)=\prod_{\alpha\in\Delta} \Upsilon^{\mathcal{B}_{\alpha(\mathfrak{m})}}\Big(\ii\alpha(a)+\frac{\R_3}{\X_3}\alpha(\mathfrak{m})\Big|\epsilon_1,\epsilon_2\Big),
\end{equation}
where $\Upsilon^{\mathcal{B}}$ is a modified version of the generalised $\Upsilon$-function:
\begin{equation}
    \Upsilon^{\mathcal{B}}(x|\epsilon_1,\epsilon_2)=\prod_{(n_1,n_2)\in\mathcal{B}}(n_1\epsilon_1+n_2\epsilon_2+x)\prod_{(n_1,n_2)\in\mathring{\mathcal{B}}}(n_1\epsilon_1+n_2\epsilon_2+\bar x).
\end{equation}
Here
\begin{equation}\label{eq.equivariance}
    \epsilon_1:=\R_1-\frac{\X_1}{\X_3}\R_3,\qquad\epsilon_2:=\R_2-\frac{\X_2}{\X_3}\R_3
\end{equation}
(assuming $\X_3\neq0$, otherwise pick $\X_1$ or $\X_2$ correspondingly) and 
\begin{equation}
    \mathcal{B}_{\alpha(\mathfrak{m})}=\{(n_1,n_2)\in\mathbb{Z}^2\,|\,(n_1,n_2,\frac{1}{\X_3}(t-n_1\X_1-n_2\X_2))\in\mathcal{C}_{\alpha(\mathfrak{m})}\}
\end{equation}
is a slice of the moment map cone $\mathcal{C}$.

\paragraph{Instanton Part.}
Dimensional reduction of the instanton part of the 5d theory on $M$ was performed in \cite{Festuccia:2016gul,Festuccia:2019akm}. Briefly, one notes that the closed Reeb orbits (around which $\mathbb{C}^2\times S^1$ is a neighbourhood) at each vertex of the moment polytope precisely agree with the $S^1$-fibres generated by $\X$. Thus, upon dimensional reduction one simply obtains Nekrasov's partition function on $\mathbb{C}^2$ around each fixed point of $v$ (the descendant of $\R$). However, at vertices where $\R$ and $\X$ are anti-parallel, in 4d we obtain anti-instantons\footnote{Briefly, this is because (horizontal) self-and anti-self-duality for 5d connections is defined with respect to the orientation transverse to $\R$, $\iota_\R\vol_M$, while in 4d it is defined with respect to the orientation transverse to $\X$, $\iota_\X\vol_M$.}. The instanton counting parameter at each vertex is determined by $\beta_\ell$ \eqref{eq.equivariance.loc} as follows\footnote{Dimensionally reducing the 5d action \eqref{eq.action5d}, no $\theta$-term arises. We can either set it to zero or add it by hand in 4d as in \cite{Festuccia:2016gul}.}:
\begin{equation}
    q_\ell=\exp\left(-16\pi^3\frac{\beta_\ell}{g^2_{5d}}\right)=\exp(2\pi\ii\tau_\ell),
\end{equation}
where $\tau_\ell=\frac{4\pi\ii}{\tilde{g}^2_{4d,\ell}}$. Finally, due to the existence of non-trivial flat connections in $\mathcal{T}_{M/\mathbb{Z}_h}$ and, correspondingly, flux solutions in $\mathcal{T}_B$, the Coulomb branch parameter $a$ receives a shift identical to the one observed in the one-loop part (after factorisation, which we do not perform here; see \cite{Lundin:2021zeb,Lundin:2023tzw}) proportional to $\mathfrak{m}$. 

In conclusion, for a theory $\mathcal{T}_B$ localising to instantons at the first $p$ fixed points and anti-instantons at the remaining $q$ ones, the instanton part reads
\begin{equation}\label{eq.inst.B}
    Z_B^\text{inst}=\prod_{\ell=1}^pZ^\text{inst}_{\mathbb{C}^2}(\ii a+\beta^{-1}_\ell\mathfrak{m}|q_\ell,\epsilon^i_\ell,\epsilon^\ell_2)\prod_{\ell=p+1}^{p+q}Z^\text{anti-inst}_{\mathbb{C}^2}(\ii a+\beta^{-1}_\ell\mathfrak{m}|\bar q_\ell,\epsilon_1^\ell,\epsilon_2^\ell).
\end{equation}
Note that the position-dependence of the complexified coupling $\tau$ above is absent in the corresponding 4d theories studied in \cite{Festuccia:2018rew,Festuccia:2019akm}. This is similar to what we observed for the classical part and will be discussed further in the next section.
    
\paragraph{Partition Function on the Base.}
Collecting the classical contribution \eqref{eq.classical.4d}, the one-loop determinant \eqref{eq.1loop.B} and the instantons \eqref{eq.inst.B},  we can write down the partition function for an $\mathcal{N}=2$ pure gauge theory on $B$, both for topologically twisted and Pestun-like theories\footnote{Recall that, if $M$ has non-trivial two-cycles it admits gauge configurations with flux which contribute also on $B$ \cite{Ruggeri:2025kmk}. We neglect them in this work as we will soon restrict to $S^5$.}
\begin{equation}
    \mathcal{Z}[\mathcal{T}_B]=\sum_{\mathfrak{m}}\int_{\ii\mathfrak{t}}\dd a\,Z_B^\text{cl}\cdot Z_B^\text{1-loop}\cdot Z_B^\text{inst}.
\end{equation}
This is the starting point for computing the sum over residues for an $SU(2)$ topologically twisted theory on $\mathbb{CP}^2$.

\section{Analytic Structure of the Partition Function on $\mathbb{CP}^2$}\label{sec.3}
In the remainder of this work we perform the integration for the partition function of a topologically twisted theory on $\mathbb{CP}^2$. We start, in this section, by studying the distribution of poles of the full partition function. We also demonstrate that all poles contributing to the residue sum can be related to contributions in \cite{Bershtein:2015xfa,Bonelli:2020xps}, where the partition function is expressed in terms of a triplet of equivariant fluxes, one for each toric divisor of $\mathbb{CP}^2$. 

Our focus will be on the $SU(2)$ gauge theory, but we will also comment on how to generalize to $SU(N)$. For $SU(2)$ we pick the positive root to be $\alpha=(1,-1)$, such that $\alpha(a)=2a$. Similarly, $\alpha(\mathfrak{m})=2\mathfrak{m}_1$. As we will also discuss certain aspects of the $SU(3)$ theory, let us define the positive roots to be
\begin{equation}
    \alpha_1=(1,-1,0),\quad\alpha_2=(1,0,-1),\quad\alpha_3=(0,1,-1).
\end{equation}
We also take $a=\diag(a_1,a_2-a_1,-a_2)$ and $\mathfrak{m}=\diag(\mathfrak{m}_1,\mathfrak{m}_2-\mathfrak{m}_1,-\mathfrak{m}_2)$.

\subsection{Partition Function on $\mathbb{CP}^2$}
As we obtain the result via the dimensional reduction reviewed in \autoref{sec.2}, the starting point is $M=S^5$. This setup has been originally treated in \cite{Lundin:2021zeb}, to which we refer for more details (see also \cite{Mauch:2025kqb} for a nice review).

\paragraph{Geometry.}
The integer points in the moment map cone $C(S^5)=\mathbb{C}^3$, that enter the perturbative partition function \eqref{eq.1loop.M}, are 
\begin{equation}
    \mathcal{C}\cap\mathbb{Z}^3=\{(n_1,n_2,n_3)\in\mathbb{Z}^3_{\geq 0}\}.
\end{equation}
We denote by $e_i$, $i=1,2,3$ the vector fields on $S^5$ generating the standard $T^3$-action and take $\X=e_1+e_2+e_3$, which coincides with the Reeb vector field for the standard choice of contact form on $S^5$. Correspondingly, the charge for the rotation along the fiber is
\begin{equation}
    t=n_1+n_2+n_3\geq 0.
\end{equation}
The fibration determined by $\X$ is precisely the Hopf-fibration with base $\mathbb{CP}^2$. Moreover, for our choice of $\X$, all signs in \eqref{eq.freeS1} are ``$+$'' (note that the $v_i$ are simply the canonical unit vectors in $\mathbb{R}^3$ here), hence, $\mathcal{T}_{\mathbb{CP}^2}$ corresponds to the topological twist. In order to retain equivariance in $\mathcal{T}_{\mathbb{CP}^2}$ after the reduction, we squash the five sphere. This has the consequence of deforming the contact form slightly compared to the standard one. Hence, we choose\footnote{We denote $\omega_i$ the entries of the Reeb vector field on $S^5$, instead of $\R_i$, to follow the standard notation in the literature.} $\R=\omega_1e_1+\omega_2e_2+\omega_3e_3$ for the Reeb vector field, where $|\omega_i-1|\ll1$. The $\omega_i$ are denoted squashing parameters.

Finally, the local equivariance parameters \eqref{eq.equivariance.loc} at the three fixed points of $\mathbb{CP}^2$ are 
\begin{center}
    \begin{tabular}{ c | c | c | c }
     & $\ell=1$ & $\ell=2$ & $\ell=3$  \\ \hline
    $\epsilon^\ell_1$ & $\epsilon_1$ & $\epsilon_2-\epsilon_1$ & $-\epsilon_2$   \\ \hline
    $\epsilon^\ell_2$ &  $\epsilon_2 $ & $ -\epsilon_1$ & $ \epsilon_1-\epsilon_2 $ 
    \end{tabular}
\end{center}
It will be useful to rewrite the squashing parameters $\omega_i$ as follows
\begin{equation}\label{eq.epsrewrite}
    \omega_1=1-\frac{\epsilon_1+\epsilon_2}{3},\quad\omega_2=1+\frac{2\epsilon_1-\epsilon_2}{3},\quad\omega_3=1+\frac{2\epsilon_2-\epsilon_1}{3}.
\end{equation}
In deriving these expressions we used the definition of the equivariance parameters \eqref{eq.equivariance} and we imposed the squashing to act on the base only, which sets $\omega_1+\omega_2+\omega_3=3$.

\paragraph{Product Over Roots and Sum Over Fluxes.}
Upon dimensional reduction, the projection condition \eqref{eq.projection.B} for the $SU(N)$ gauge theory sets
\begin{equation}\label{eq.projection.CP2}
    t=\alpha(\mathfrak{m})\geq 0,
\end{equation}
where recall that $\mathfrak{m}_i\geq 0$ as they arise from winding numbers of flat connections. These conditions constrain both the product over roots in the one-loop determinant and the sum over fluxes in the full partition function. How the former product is affected is quite simple: we set the product over roots in \eqref{eq.1loop.B} to be over positive roots $\Delta_+$ only.

Understanding how the sum over fluxes is affected is more insightful. First, we consider $G=SU(2)$. In this case the projection condition $\alpha(\mathfrak{m})\geq 0$ does not further restrict the possible values of $\mathfrak{m}_1$. However, it becomes more interesting for higher-rank gauge groups. Let us take $SU(3)$. Then, the projection condition sets
\begin{equation}\label{eq.proj.SU3}
    2\mathfrak{m}_1\geq \mathfrak{m}_2,\quad \mathfrak{m}_1+\mathfrak{m}_2\geq 0,\quad 2\mathfrak{m}_2\geq\mathfrak{m}_1.
\end{equation}
Hence, the sum over fluxes needs to be restricted accordingly. At the end of this section, we will compare these restriction to the stability conditions of gauge bundles over $\mathbb{CP}^2$ in \cite{Bershtein:2015xfa,Bonelli:2020xps}. The same conditions can be studied for $SU(N)$. We then introduce
\begin{equation}
    \Delta_\mathfrak{m}=\{\mathfrak{m}\in\mathbb{N}^{N\times N}\,|\,\alpha(\mathfrak{m})\geq 0,\,\forall\alpha\in\Delta_+\},
\end{equation}
which labels the set of flux sectors which satisfy the projection condition.

\paragraph{Partition Function.}
The full partition function for an $SU(2)$ topologically twisted theory on $\mathbb{CP}^2$ is
\begin{equation}\label{eq.ZCP2}
    \mathcal{Z}[\mathcal{T}_{\mathbb{CP}^2}]=\sum_{\mathfrak{m}\in\Delta_\mathfrak{m}}\int_{\ii\mathbb{R}}\dd a\,Z_{\mathbb{CP}^2}^\text{cl}\cdot Z_{\mathbb{CP}^2}^\text{1-loop}\cdot Z_{\mathbb{CP}^2}^\text{inst}.
\end{equation}
The integration contour is along the imaginary axis due to the Wick rotation discussed in \cite{Qiu:2016dyj}. Moreover, the four dimensional position-dependent coupling \eqref{eq.coupling} is given, at each fixed point, by
\begin{equation}
    \tilde{g}^2_{4d,\ell}=\frac{g^2_{5d}h}{2\pi}\omega_\ell=g^2_\text{4d}\omega_\ell.
\end{equation}
Hence, the classical part, $Z_{\mathbb{CP}^2}^\text{cl}$, evaluates to
\begin{equation}\label{eq.class.CP2}
    Z_{\mathbb{CP}^2}^\text{cl}=\exp\left(\sum_{\ell=1}^3\frac{(2\pi)^3}{g^2_{\text{4d}}\omega_\ell}\frac{\tr (\ii a+\omega_\ell\mathfrak{m})^2}{\epsilon_1^\ell\epsilon_2^\ell}\right)=\exp\left(-\frac{(2\pi)^2}{g_\text{4d}^2}\frac{\tr (a^2)}{\omega_1\omega_2\omega_3}\right)=q^{\frac{\tr(a)^2}{2\omega_1\omega_2\omega_3}}.
\end{equation}
Note that the absence of flux-terms (i.e. terms containing $\mathfrak{m}$) is due to the position-dependence of the 4d Yang-Mills coupling. If the latter was not present, we would instead obtain $\exp(-(2\pi)^2\mathfrak{m}^2/g_\text{4d}^2)$. This comment will be relevant at the end of this section when comparing with the classical contribution in \cite{Bershtein:2015xfa,Bonelli:2020xps}.

\subsection{Zeroes and Poles}
We now study the zeroes and poles in the partition function \eqref{eq.ZCP2}. Hence, we rewrite the Coulomb branch parameter $a$ as
\begin{equation}\label{eq.poles.US}
    \hat{a}=\text{Re}(a)+\ii\mathfrak{m}+\frac{\ii}{2}\left(\left(k-\frac{2}{3}\mathfrak{m}\right)\epsilon_1+\left(l-\frac{2}{3}\mathfrak{m}\right)\epsilon_2\right).
\end{equation}
The shifts proportional to $\mathfrak{m}$ are introduced to facilitate the comparison with \cite{Bershtein:2015xfa,Bonelli:2020xps}. All poles and zeroes will be along the imaginary line and thus we set $\text{Re}(a)=0$. We can then rewrite the partition function as a sum over residues\footnote{We restrict the sum over strictly positive $\mathfrak{m}_1$. This is required to obtain the correct prefactor in front of semi-stable contributions to be discussed shortly.}
\begin{equation}\label{eq.sumresidue}
    \mathcal{Z}[\mathcal{T}_{\mathbb{CP}^2}]=\sum_{\mathfrak{m}_1> 0}\sum_{j,k}\text{Res}_{a=\hat{a}}\,Z_{\mathbb{CP}^2}^\text{full},
\end{equation}
where we defined 
\begin{equation}
    Z_{\mathbb{CP}^2}^\text{full}=Z_{\mathbb{CP}^2}^\text{cl}\cdot Z_{\mathbb{CP}^2}^\text{1-loop}\cdot Z_{\mathbb{CP}^2}^\text{inst}.
\end{equation}

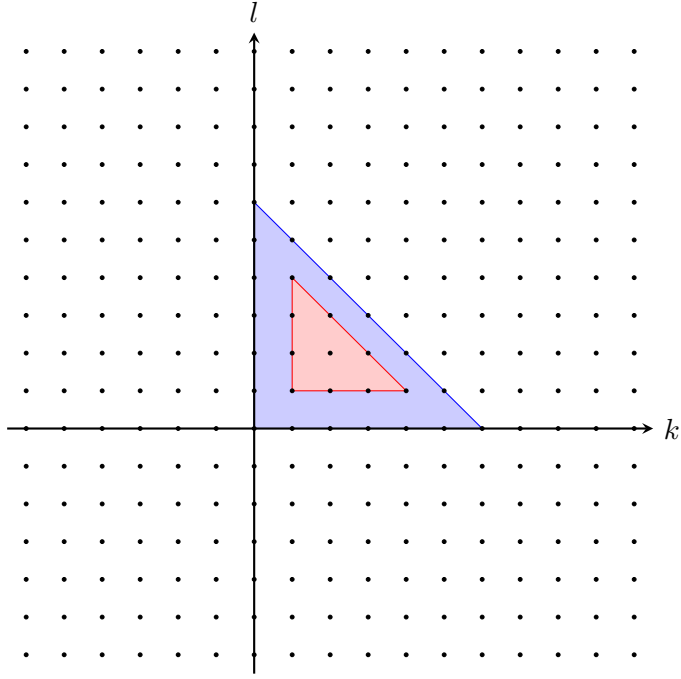
\begin{figure}[ht]
\centering
\begin{tikzpicture}[scale=.5]
\def\rad{.13cm}
\def\rrad{.05cm}
\filldraw[draw=blue,fill=blue!20,opacity=1]       
(-2,-2)
-- (-2,4)
-- (4,-2)
-- cycle;
\filldraw[draw=red,fill=red!20,opacity=1]       
(-1,-1)
-- (-1,2)
-- (2,-1)
-- cycle;
\draw[-stealth,thick] (-8.5,-2)--(8.5,-2) node [right]{$k$};
\draw[-stealth,thick] (-2,-8.5)--(-2,8.5) node [above]{$l$};
\foreach \n in {-8,...,8}{
\foreach \m in {-8,...,8}{
\draw[fill=black] (\n,\m) circle [radius=\rrad];}}
\end{tikzpicture}
\caption{Zeroes distribution of the one-loop determinant for $\mathfrak{m}_1=3$. Simple zeroes are in the blue region and double zeroes are in the red region. The remaining points are regular points.}
\label{fig.1loop}
\end{figure}

\paragraph{One-Loop Determinant.}
We start by analysing the poles and zeroes of the one-loop determinant \eqref{eq.1loop.B} at a given flux sector $\mathfrak{m}_1$. This only contributes with zeroes, and it is given by
\begin{equation}\begin{split}\label{eq.1loopCP2}
     Z_{\mathbb{CP}^2}^\text{1-loop}
    =\prod_{(n_1,n_2)\in\mathcal{B}_{\mathfrak{m}_1}}\left(\epsilon_1n_1+\epsilon_2n_2+2\ii a+2\left(1-\frac{\epsilon_1+\epsilon_2}{3}\right)\mathfrak{m}_1\right) \\
    \prod_{(n_1,n_2)\in\mathcal{B}^\circ_{\mathfrak{m}_1}}\left(\epsilon_1n_1+\epsilon_2n_2+2\ii a +2\left(1-\frac{\epsilon_1+\epsilon_2}{3}\right)\mathfrak{m}_1\right),
\end{split}\end{equation}
where
\begin{equation}\label{eq.Bm}
    \mathcal{B}_{\mathfrak{m}_1}=\{(n_1,n_2)\in\mathbb{Z}^2_{\geq 0}\,|\,n_1+n_2\leq 2\mathfrak{m}_1\}.
\end{equation}
Evaluating this expression on \eqref{eq.poles.US} we find
\begin{equation}\label{eq.1loopCP2res}
    Z_{\mathbb{CP}^2}^\text{1-loop}\big|_{a=\hat{a}}=\prod_{(n_1,n_2)\in\mathcal{B}_{\mathfrak{m}_1}}\left(\epsilon_1(n_1-k)+\epsilon_2(n_2-j)\right)\prod_{(n_1,n_2)\in\mathcal{B}^\circ_{\mathfrak{m}_1}}\left(\epsilon_1(n_1-k)+\epsilon_2(n_2-l)\right),
\end{equation}
from which one can read the values of the zeroes, as show in \autoref{fig.1loop}.
As we are integrating over $a\in\ii\mathbb{R}$ we find that all zeroes are along the integration contour.

\paragraph{Instantons.}
The instanton partition function for the topologically twisted theory on $\mathbb{CP}^2$ is given by three Nekrasov partition function, one for each fixed point \eqref{eq.inst.B}. A single contribution for $G=SU(2)$ is written as a sum over the array $\vec{Y}=(Y_1,Y_2)$ of two Young diagrams
\begin{equation}
    Z^\text{inst}_{\mathbb{C}^2}(a|q,\epsilon_1,\epsilon_2)=\sum_{\vec{Y}} q^{|\vec{Y}|}Z_{\vec{Y}}(a|\epsilon_1,\epsilon_2),
\end{equation}
where
\begin{equation}\begin{split}
    Z_{\vec{Y}}(a|\epsilon_1,\epsilon_2)=\prod_{u,v=1}^2&\prod_{(n_1,n_2)\in Y_u}\left(2a+\epsilon_1(n_1-l_{vn_2})-\epsilon_2(n_2-1-\tilde{l}_{un_1})\right)^{-1}\\
    &\prod_{(n_1,n_2)\in Y_v}\left(2a-\epsilon_1(n_1-1-l_{un_2})+\epsilon_2(n_2-\tilde{l}_{vn_1})\right)^{-1}
\end{split}\end{equation}
and $\{l_{un_2}\}$, $\{\tilde{l}_{vn_1}\}$ denote, respectively, the length of the rows and columns of $Y_u$.

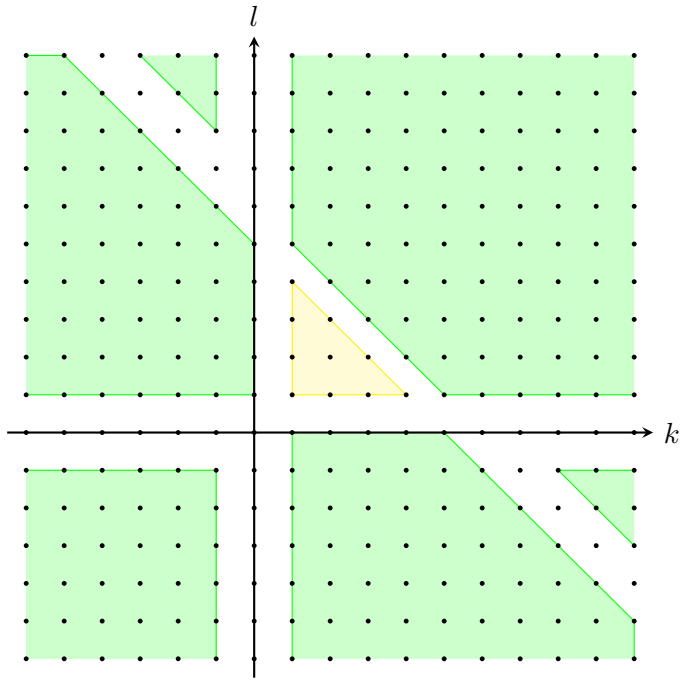
\begin{figure}[ht]
\centering
\begin{tikzpicture}[scale=.5]
\def\rad{.13cm}
\def\rrad{.05cm}
\filldraw[draw=yellow,fill=yellow!20,opacity=1]       
(-1,-1)
-- (-1,2)
-- (2,-1)
-- cycle;
\fill[fill=green!20,opacity=1]       
(-3,-3)
-- (-3,-8)
-- (-8,-8)
-- (-8,-3)
-- cycle;
\draw[draw=green,opacity=1]       
(-8,-3)
-- (-3,-3)
-- (-3,-8);
\fill[fill=green!20,opacity=1]       
(-2,-1)
-- (-8,-1)
-- (-8,8)
-- (-7,8)
-- (-2,3)
-- cycle;
\draw[draw=green,opacity=1]       
(-8,8)
-- (-7,8)
-- (-2,3)
-- (-2,-1)
-- (-8,-1);
\fill[fill=green!20,opacity=1]       
(-3,6)
-- (-5,8)
-- (-3,8)
-- cycle;
\draw[draw=green,opacity=1]       
 (-3,8)
-- (-3,6)
-- (-5,8);
\fill[fill=green!20,opacity=1]       
(-1,8)
-- (-1,3)
-- (3,-1)
-- (8,-1)
-- (8,8)
-- cycle;
\draw[draw=green,opacity=1]       
(-1,8)
-- (-1,3)
-- (3,-1)
-- (8,-1);
\fill[fill=green!20,opacity=1]       
(6,-3)
-- (8,-5)
-- (8,-3)
-- cycle;
\draw[draw=green,opacity=1]       
(8,-3)
-- (6,-3)
-- (8,-5);
\fill[fill=green!20,opacity=1]       
(-1,-2)
-- (-1,-8)
-- (8,-8)
-- (8,-7)
-- (3,-2)
-- cycle;
\draw[draw=green,opacity=1]       
(8,-8)
-- (8,-7)
-- (3,-2)
-- (-1,-2)
-- (-1,-8);
\draw[-stealth,thick] (-8.5,-2)--(8.5,-2) node [right]{$k$};
\draw[-stealth,thick] (-2,-8.5)--(-2,8.5) node [above]{$l$};
\foreach \n in {-8,...,8}{
\foreach \m in {-8,...,8}{
\draw[fill=black] (\n,\m) circle [radius=\rrad];}}
\end{tikzpicture}
\caption{Poles distribution of the instanton partition function for $\mathfrak{m}_1=3$. Simple poles are in the green region and triple poles are in the yellow region. The remaining points are regular points.}
\label{fig.inst}
\end{figure}

For the $SU(2)$ theory, the Nekrasov partition function on $\mathbb{C}^2$ can be expressed using Zamolodchikov's form \cite{Poghossian:2009mk}. If we combine the three fixed points contribution of $\mathbb{CP}^2$ as in \eqref{eq.inst.B}, we obtain the following\footnote{This rewriting can be generalized to $SU(N)$ gauge theories \cite{Sysoeva:2022syp}.}
\begin{equation}\begin{split}\label{eq.instCP2}
    Z_{\mathbb{CP}^2}^\text{inst}=&\left(1-\sum_{n_1,n_2=1}^\infty\frac{q^{n_1n_2}_1R^{1}_{n_1n_2}Z_{\mathbb{C}^2}^\text{inst}(n_1\epsilon_1-n_2\epsilon_2|q_1,\epsilon_1,\epsilon_2)}{(2\ii a^{1}-n_1\epsilon_1-n_2\epsilon_2)(2\ii a^{1}+n_1\epsilon_1+n_2\epsilon_2)}\right)\\
    &\left(1-\sum_{n_2,n_3=1}^\infty\frac{q^{n_2n_3}_2R^{2}_{n_2n_3}Z_{\mathbb{C}^2}^\text{inst}(n_2(\epsilon_2-\epsilon_1)-n_3(-\epsilon_1)|q_2,\epsilon_2-\epsilon_1,-\epsilon_1)}{(2\ii a^{2}-n_2(\epsilon_2-\epsilon_1)-n_3(-\epsilon_1))(2\ii a^{2}+n_2(\epsilon_2-\epsilon_1)+n_3(-\epsilon_1))}\right)\\
    &\left(1-\sum_{n_3,n_1=1}^\infty\frac{q^{n_3n_1}_3R^{3}_{n_3n_1}Z_{\mathbb{C}^2}^\text{inst}(n_3(-\epsilon_2)-n_1(\epsilon_1-\epsilon_2)|q_3,-\epsilon_2,\epsilon_1-\epsilon_2)}{(2\ii a^{3}-n_3(-\epsilon_2)-n_1(\epsilon_1-\epsilon_2))(2\ii a^{3}+n_3(-\epsilon_2)+n_1(\epsilon_1-\epsilon_2))}\right),
\end{split}\end{equation}
where the local Coulomb branch parameter is
\begin{equation}\begin{split}
    \ii a^{1}=&\ii a+\left(1-\frac{\epsilon_1+\epsilon_2}{3}\right)\mathfrak{m}_1,\\
    \ii a^{2}=&\ii a+\left(1+\frac{2\epsilon_1-\epsilon_2}{3}\right)\mathfrak{m}_1,\\
    \ii a^{3}=&\ii a+\left(1+\frac{2\epsilon_2-\epsilon_1}{3}\right)\mathfrak{m}_1, 
\end{split}\end{equation}
and
\begin{equation}\label{eq.Rn1n2}
    R_{n_\ell n_{\ell+1}}^{\ell}=2\prod_{i=-n_\ell-1}^{n_\ell}\prod_{j=n_{\ell+1}-1\atop (i,j)\neq (0,0),(n_\ell,n_{\ell+1})}^{n_{\ell+1}}\frac{1}{i\epsilon_1^\ell+j\epsilon_2^\ell}.
\end{equation}
As for the one-loop part, we rewrite \eqref{eq.instCP2} substituting \eqref{eq.poles.US}
\begin{equation}\begin{split}\label{eq.instCP2res}
    &Z_{\mathbb{CP}^2}^\text{inst}\big|_{a=\hat{a}}=\left(1-\sum_{n_1,n_2=1}^\infty\frac{q^{n_1n_2}_1R^{1}_{n_1n_2}Z^\text{inst}(n_1\epsilon_1+n_2\epsilon_2,\epsilon_1,-\epsilon_2,q_1)}{(-(n_1+k)\epsilon_1-(n_2+l)\epsilon_2)((n_1-k)\epsilon_1+(n_2-l)\epsilon_2)}\right)\\
    &\left(1-\sum_{n_2,n_3=1}^\infty\frac{q^{n_2n_3}_2R^{2}_{n_2n_3}Z^\text{inst}(n_2(\epsilon_2-\epsilon_1)-n_3(-\epsilon_1),\epsilon_2-\epsilon_1,-\epsilon_1,q_2)}{(-(n_2+l)(\epsilon_2-\epsilon_1)-(n_3+p)(-\epsilon_1))((n_2-l)(\epsilon_2-\epsilon_1)+(n_3-p)(-\epsilon_1))}\right)\\
    &\left(1-\sum_{n_3,n_1=1}^\infty\frac{q^{n_3n_1}_3R^{3}_{n_3n_1}Z^\text{inst}(n_3(-\epsilon_2)-n_1(\epsilon_1-\epsilon_2),-\epsilon_2,\epsilon_1-\epsilon_2,q_3)}{(-(n_3+p)(-\epsilon_2)-(n_1+k)(\epsilon_1-\epsilon_2))((n_3-p)(-\epsilon_2)+(n_1-k)(\epsilon_1-\epsilon_2))}\right),
\end{split}\end{equation}
where we renamed the combination
\begin{equation}\label{eq.p}
    2\mathfrak{m}-k-l\equiv p.
\end{equation}
The poles of the instanton contribution \eqref{eq.instCP2} are shown in \autoref{fig.inst}.
All poles lie along the integration contour of $a$.

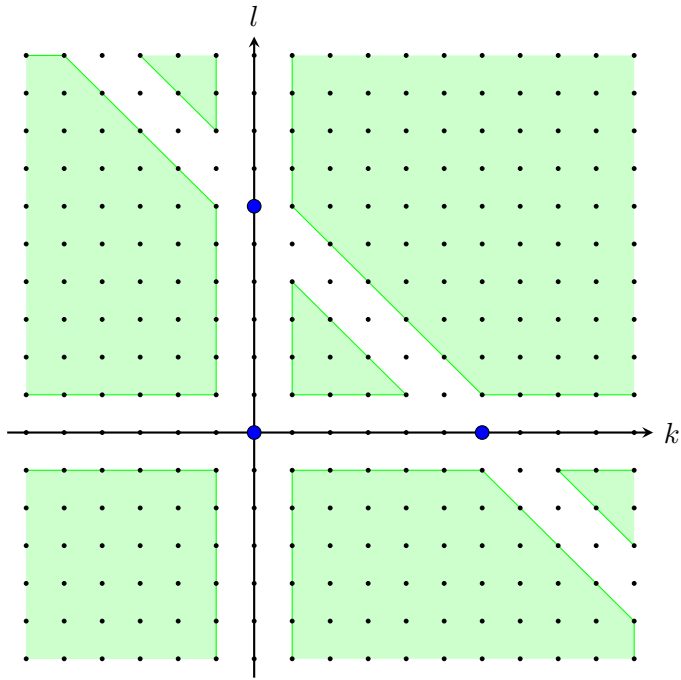
\begin{figure}[ht]
\centering
\begin{tikzpicture}[scale=.5]
\def\rad{.18cm}
\def\rrad{.05cm}
\filldraw[draw=green,fill=green!20,opacity=1]       
(-1,-1)
-- (-1,2)
-- (2,-1)
-- cycle;
\fill[fill=green!20,opacity=1]       
(-3,-3)
-- (-3,-8)
-- (-8,-8)
-- (-8,-3)
-- cycle;
\draw[draw=green,opacity=1]       
(-8,-3)
-- (-3,-3)
-- (-3,-8);
\fill[fill=green!20,opacity=1]       
(-3,-1)
-- (-8,-1)
-- (-8,8)
-- (-7,8)
-- (-3,4)
-- cycle;
\draw[draw=green,opacity=1]       
(-8,8)
-- (-7,8)
-- (-3,4)
-- (-3,-1)
-- (-8,-1);
\fill[fill=green!20,opacity=1]       
(-3,6)
-- (-5,8)
-- (-3,8)
-- cycle;
\draw[draw=green,opacity=1]       
 (-3,8)
-- (-3,6)
-- (-5,8);
\fill[fill=green!20,opacity=1]       
(-1,8)
-- (-1,4)
-- (4,-1)
-- (8,-1)
-- (8,8)
-- cycle;
\draw[draw=green,opacity=1]       
(-1,8)
-- (-1,4)
-- (4,-1)
-- (8,-1);
\fill[fill=green!20,opacity=1]       
(6,-3)
-- (8,-5)
-- (8,-3)
-- cycle;
\draw[draw=green,opacity=1]       
(8,-3)
-- (6,-3)
-- (8,-5);
\fill[fill=green!20,opacity=1]       
(-1,-3)
-- (-1,-8)
-- (8,-8)
-- (8,-7)
-- (4,-3)
-- cycle;
\draw[draw=green,opacity=1]       
(8,-8)
-- (8,-7)
-- (4,-3)
-- (-1,-3)
-- (-1,-8);
\draw[-stealth,thick] (-8.5,-2)--(8.5,-2) node [right]{$k$};
\draw[-stealth,thick] (-2,-8.5)--(-2,8.5) node [above]{$l$};
\foreach \n in {-8,...,8}{
\foreach \m in {-8,...,8}{
\draw[fill=black] (\n,\m) circle [radius=\rrad];}}
\draw[fill=blue] (-2,-2) circle [radius=\rad];
\draw[fill=blue] (-2,4) circle [radius=\rad];
\draw[fill=blue] (4,-2) circle [radius=\rad];
\end{tikzpicture}
\caption{Poles and zeroes distribution of the full partition function for $\mathfrak{m}_1=3$. Simple poles are in the green region and simple zeroes are the blue points. The remaining points are regular points.}
\label{fig.full}
\end{figure}

\paragraph{Full Partition Function.}
By combining \autoref{fig.1loop} and \autoref{fig.inst} we plot the poles and zeroes distribution of the full partition function in \autoref{fig.full}.
As expected from the previous analysis, all critical points of the full partition function lie along the integration contour of the Coulomb branch parameter $a$.

\subsection{Comparison with Equivariant Fluxes}
The distribution of poles and zeroes in \autoref{fig.full} shows a striking similarity with Figure 2 in \cite{Bonelli:2020xps}, where the partition function of the $\mathcal{N}=2$ $SU(2)$ topologically twisted theory on $\mathbb{CP}^2$ is computed in terms of equivariant fluxes $(k^1,k^2,k^3)$, one for each toric divisor of the manifold. The integers $k^\ell$ satisfy\footnote{We restrict the sum of equivariant fluxes to be even since that corresponds to the $SU(2)$ theory in \cite{Bershtein:2016mxz,Bonelli:2020xps}. Odd values correspond, instead, to $SO(3)$ gauge bundles.}
\begin{equation}\label{eq.kellm}
    k^1+k^2+k^3=2\mathfrak{m}_1.
\end{equation}
Here, we only present their result for the partition function and the integration contour to facilitate the comparison with our setup. We refer to \cite{Bershtein:2015xfa,Bershtein:2016mxz,Bonelli:2020xps} for more details\footnote{See also \cite{Kim:2025fpz} for recent analysis of the K-theoretic partition function on $\mathbb{CP}^2$.}. The full partition function is
\begin{equation}\label{eq.BT1}
    \mathcal{Z}[\mathcal{T}_{\mathbb{CP}^2}^\text{equiv}]=\sum_{(k^1,k^2,k^2)\in\mathbb{Z}^3\atop \mathfrak{m}_1> 0}\int_{\mathbb{C}/\{0\}}\dd a\,\dd\bar{a}\,Z_\text{equiv}^\text{cl}\cdot Z_\text{equiv}^\text{1-loop}\cdot Z_\text{equiv}^\text{inst}
\end{equation}
A first comment regards the sum over equivariant fluxes, which is restricted to $\mathfrak{m}_1> 0$. This condition ensures that only (semi-)stable $SU(2)$ bundles are summed over. This exactly matches the condition arising from the projection condition \eqref{eq.projection.CP2}. Nicely, the equivalence continues to hold for gauge groups of higher rank. Considering $SU(3)$, the stability condition reads \cite{Bonelli:2020xps}
\begin{equation}
    \mathfrak{m}_1+\mathfrak{m}_2\geq 0,\qquad 2\mathfrak{m}_1\geq\mathfrak{m}_2,
\end{equation}
where $2\mathfrak{m}_i=k^1_i+k^2_i+k^3_i$. Up to a Weyl symmetry, this reproduces \eqref{eq.proj.SU3}. Hence, we discover a nice feature of the dimensional reduction from 5d via the $\mathbb{Z}_h$ quotient: the projection condition \eqref{eq.projection} naturally encodes, at large $h$, the stability condition of gauge bundles over $\mathbb{CP}^2$. It would be interesting to extend this relation to different (quasi-)toric four-manifolds.

Considering the $SU(2)$ gauge theory, for each value of $\mathfrak{m}_1$, there is an infinite number of triples $(k^1,k^2,k^3)$ which one has to sum over in the partition function. In the partition function obtained from 5d, instead, there is only one summand for each value of $\mathfrak{m}_1$. This mismatch is compensated by a different integration contour. In \cite{Bershtein:2015xfa}, the complex scalars $(\Phi,\bar{\Phi})$, where $\bar{\Phi}=\Phi^\dagger$, evaluate on the BPS locus to
\begin{equation}
    \Phi=a+\sum_{\ell=1}^3 k^\ell H_\ell,
\end{equation}
where $H_\ell$ is the zero-form part of a $T^2$-equivariant two-form which is the Poincaré dual of an equivariant divisor. The integral in \eqref{eq.BT1} is then over zero-modes for both components of the $\mathcal{N}=2$ complex scalar, unlike in our case where one of the two components is set to be proportional to the gauge flux\footnote{The authors of \cite{Festuccia:2018rew} also commented about a discrepancy between their BPS solution and the one in \cite{Bershtein:2015xfa}. In particular they find that $F+\iota_\R b\,\varphi_4$ is not the curvature of an equivariant line bundle. As our solution reproduces that in \cite{Festuccia:2018rew}, this holds also in our case.} \eqref{eq.varphibps}. The integration in \eqref{eq.BT1} over $\dd a\,\dd\bar{a}$ can be shown to simplify, via Stokes theorem, to a single contribution at $a=0$
\begin{equation}\label{eq.sumresidue.BT}
    \mathcal{Z}[\mathcal{T}_{\mathbb{CP}^2}^\text{equiv}]=\sum_{(k^1,k^2,k^2)\in\mathbb{Z}^3\atop \mathfrak{m}_1>0}\text{Res}_{a=0}\,Z_\text{equiv}^\text{full},
\end{equation}
where
\begin{equation}
    Z_\text{equiv}^\text{full}=Z_\text{equiv}^\text{cl}\cdot Z_\text{equiv}^\text{1-loop}\cdot Z_\text{equiv}^\text{inst}.
\end{equation}
Hence, for each triple of equivariant fluxes at most one point, if it is a pole, is included in the residue sum.

Comparing the one-loop determinant and instanton contribution in \cite{Bershtein:2015xfa}, at $a=0$, with \eqref{eq.1loopCP2res} and \eqref{eq.instCP2res} at $a=\hat{a}$ \eqref{eq.poles.US}, it is immediate to check that they do match up to the relabelling 
\begin{equation}\label{eq.BTUSmap1}
    \left(k^1,k^2,k^3\right)\;\leftrightarrow\;\left(k,l,p\right).
\end{equation}
Hence, under this map, \autoref{fig.full} shows all the residues in the $(k^1,k^2)$-plane, contributing to \eqref{eq.sumresidue.BT}, and arising from all triples $(k^1,k^2,k^3)$ such that $k^1+k^2+k^3=6$. 

Treating the classical piece requires more care. In \cite{Bershtein:2015xfa} this is given by
\begin{equation}\label{eq.classBT}
    Z^\text{cl}_\text{equiv}=\sum_{\ell=1}^3\frac{(2\pi)^2}{g^2_{\text{4d}}}\frac{\tr (a+k^\ell\epsilon_1^\ell+k^{\ell+1}\epsilon_2^{\ell+1})^2}{\epsilon_1^\ell\epsilon_2^\ell}=\exp\left(-\frac{(2\pi)^2}{g_\text{4d}^2}\frac{\mathfrak{m}_1^2}{\omega_1\omega_2\omega_3}\right),
\end{equation}
where the coupling constant is not the position dependent $\tilde{g}^2_{\text{4d}}$. This clearly does not match \eqref{eq.class.CP2}. However, if we isolate the contribution from a single patch, say $\ell=1$, in \eqref{eq.class.CP2} and we rewrite it substituting \eqref{eq.poles.US} and \eqref{eq.BTUSmap1}, we find\footnote{To derive this relation we rewrite $\omega_1$ in front of $\mathfrak{m}_1$ in terms of equivariance parameters \eqref{eq.epsrewrite}.}:
\begin{equation}\label{eq.class.comp}
    Z^{\text{cl},\ell=1}_{\mathbb{CP}^2}|_{a=\hat{a}}=\exp\left(\frac{(2\pi)^3}{\tilde{g}^2_{\text{4d},1}}\frac{2(\ii a+\omega_1\mathfrak{m}_1)^2}{\epsilon_1\epsilon_2}\right)=\exp\left(\frac{(2\pi)^2}{g^2_{\text{4d}}\omega_1}\frac{(k^1\epsilon_1+k^2\epsilon_2)^2}{\epsilon_1\epsilon_2}\right).
\end{equation}
This reproduces the contribution from the first fixed point in \eqref{eq.classBT}, evaluated at $a=0$, up to the different value of the 4d coupling constant which, via dimensional reduction, also includes a factor of $\omega_1$
\begin{equation}
    Z^{\text{cl},\ell=1}_{\mathbb{CP}^2}|_{(a=\hat{a},\, \omega_1=1)}=Z^{\text{cl},\ell=1}_\text{equiv}|_{a=0}.
\end{equation}
The same can be shown to hold for the other two fixed points. 

As noted above, when one combines all fixed point contributions, due to the different coupling constant in each patch, the resulting classical contributions are different. However, if one considers the non-equivariant limit\footnote{The non-equivariant limit also implies $\omega_i\rightarrow 1$, as one can find from \eqref{eq.equivariance}.}, $\epsilon_1,\epsilon_2\rightarrow 0$, evaluating the classical contribution \eqref{eq.class.CP2} at $\hat{a}=\ii\mathfrak{m}_1$, one finds exactly \eqref{eq.classBT}. Therefore, in this limit, the two observables precisely match. It was observed in \cite{Festuccia:2016gul} that a 4d coupling not-position dependent can be achieved at the cost of introducing a position-dependent $\theta$-term. It would be interesting to study a 5d observable that, upon dimensional reduction, directly gives rise to 4d coupling not position dependent.  A possibility would be to consider the mixed Chern-Simons terms in \cite{Baulieu:1997nj} which are not gauge-invariant.

\section{Residue Sum and Donaldson Invariants}\label{sec.4}
In this section we explicitly perform the residue sum over the poles analysed in \autoref{sec.3}. The naive contour of integration over $a$ is along the imaginary line, where all poles of the integrand lie. Therefore, it needs to be deformed slightly. We make the canonical choice to include all of the poles by running the contour along $\ii\mathbb{R}-\epsilon$ and closing it at $+\infty$. 

It would be a long but straightforward computation to sum over all residues at all flux sectors. Luckily, we can simplify the residue sum by studying the cancellations occurring between different contributions. Such simplifications can be traced back to the relation between the poles in our partition function and those in \cite{Bershtein:2015xfa,Bonelli:2020xps}. Hence, we first study these cancellations and then, at the end of the section, we perform explicit computations. This leads to new invariants which, in the non-equivariant limit, reproduce Donaldson invariants.

\subsection{Abstruse Duality}
We briefly mentioned already that many cancellations occur between different contributions in the residue sum. To understand how such cancellations arise, we isolate a single contribution from a fixed point of the integrand of the full partition function \eqref{eq.BT1} in \cite{Bershtein:2015xfa,Bonelli:2020xps}\footnote{We have already shown in \eqref{eq.class.CP2} and \eqref{eq.instCP2} that the classical and instanton parts can be factorized into contributions from the three fixed points. The same applies to the one-loop determinant \cite{Lundin:2023tzw}.}
\begin{equation}\label{eq.fullC2}
    \mathcal{Z}_{\mathbb{C}^2}^\ell(a^\ell_\text{equiv}|q,\epsilon_1^\ell,\epsilon_2^\ell)=Z_{\mathbb{C}^2}^\text{cl}\cdot Z_{\mathbb{C}^2}^\text{1-loop}\cdot Z_{\mathbb{C}^2}^\text{inst},
\end{equation}
such that
\begin{equation}
    \mathcal{Z}[\mathcal{T}_\text{equiv}^2]=\sum_{(k^1,k^2,k^2)\in\mathbb{Z}^3\atop \mathfrak{m}_1> 0}\int_{\mathbb{C}/\{0\}}\dd a\dd\bar{a}\prod_{\ell=1}^3\mathcal{Z}_{\mathbb{C}^2}^\ell(a^\ell_\text{equiv}|q,\epsilon_1^\ell,\epsilon_2^\ell).
\end{equation}
Then, the \emph{abstruse duality} holds \cite{Bonelli:2020xps}
\begin{equation}\label{eq.abstruse2}
    \lim_{a\rightarrow0}\frac{\mathcal{Z}_{\mathbb{C}^2}^\ell(a-\frac{\ii}{2}(m\epsilon_1+n\epsilon_2)|q_\ell,\epsilon_1^\ell,\epsilon_2^\ell)}{\mathcal{Z}_{\mathbb{C}^2}^\ell(a-\frac{\ii}{2}(m\epsilon_1-n\epsilon_2)|q_\ell,\epsilon_1^\ell,\epsilon_2^\ell)}=-\text{sign}(\epsilon_1^\ell).
\end{equation}
From this relation, exploiting the singularity structure of the partition function on $\mathbb{C}^2$ \eqref{eq.fullC2} and properties of the toric geometry of $\mathbb{CP}^2$, the authors of \cite{Bonelli:2020xps} have shown that 
\begin{equation}\begin{split}\label{eq.abstruse}
    \text{Res}_{a=0}\,Z_\text{equiv}^\text{full}\big|_{(k^1,k^2,k^3)}&=-\text{Res}_{a=0}\,Z_\text{equiv}^\text{full}\big|_{(-k^1,k^2,k^3)}\\
    &=-\text{Res}_{a=0}\,Z_\text{equiv}^\text{full}\big|_{(k^1,-k^2,k^3)}\\
    &=-\text{Res}_{a=0}\,Z_\text{equiv}^\text{full}\big|_{(k^1,k^2,-k^3)}
\end{split}\end{equation}
where each residue contributes to \eqref{eq.sumresidue.BT}. Despite our classical contribution on each patch being different, it only differs by the value of the coupling constant \eqref{eq.class.comp}. Hence, it is immediate to show that the abstruse duality \eqref{eq.abstruse2} continues to hold flipping signs in the triple $(k,l,p)$. Due to \eqref{eq.kellm}, flipping any of $k,l,p$ also affects the value of $\mathfrak{m}$. 

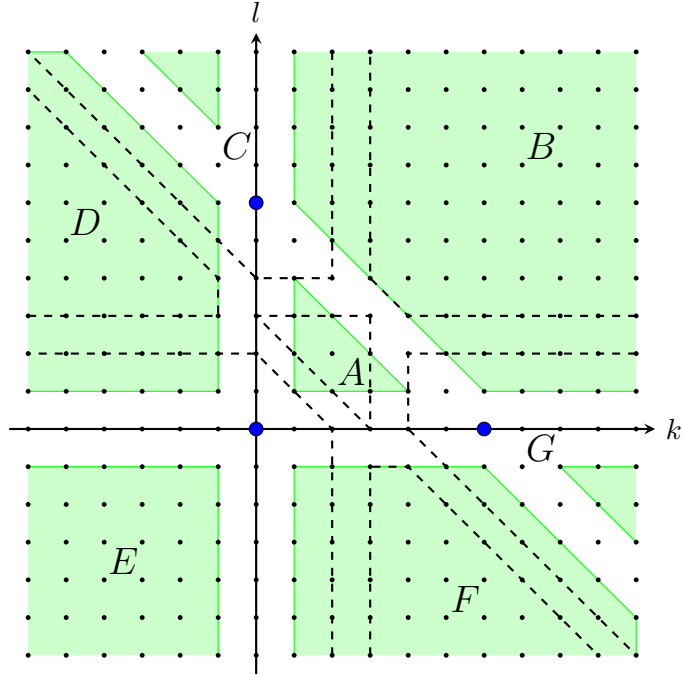
\begin{figure}[ht]
\centering
\begin{tikzpicture}[scale=.5]
\def\rad{.18cm}
\def\rrad{.05cm}
\filldraw[draw=green,fill=green!20,opacity=1]       
(-1,-1)
-- (-1,2)
-- (2,-1)
-- cycle;
\fill[fill=green!20,opacity=1]       
(-3,-3)
-- (-3,-8)
-- (-8,-8)
-- (-8,-3)
-- cycle;
\draw[draw=green,opacity=1]       
(-8,-3)
-- (-3,-3)
-- (-3,-8);
\fill[fill=green!20,opacity=1]       
(-3,-1)
-- (-8,-1)
-- (-8,8)
-- (-7,8)
-- (-3,4)
-- cycle;
\draw[draw=green,opacity=1]       
(-8,8)
-- (-7,8)
-- (-3,4)
-- (-3,-1)
-- (-8,-1);
\fill[fill=green!20,opacity=1]       
(-3,6)
-- (-5,8)
-- (-3,8)
-- cycle;
\draw[draw=green,opacity=1]       
 (-3,8)
-- (-3,6)
-- (-5,8);
\fill[fill=green!20,opacity=1]       
(-1,8)
-- (-1,4)
-- (4,-1)
-- (8,-1)
-- (8,8)
-- cycle;
\draw[draw=green,opacity=1]       
(-1,8)
-- (-1,4)
-- (4,-1)
-- (8,-1);
\fill[fill=green!20,opacity=1]       
(6,-3)
-- (8,-5)
-- (8,-3)
-- cycle;
\draw[draw=green,opacity=1]       
(8,-3)
-- (6,-3)
-- (8,-5);
\fill[fill=green!20,opacity=1]       
(-1,-3)
-- (-1,-8)
-- (8,-8)
-- (8,-7)
-- (4,-3)
-- cycle;
\draw[draw=green,opacity=1]       
(8,-8)
-- (8,-7)
-- (4,-3)
-- (-1,-3)
-- (-1,-8);
\draw[dashed,thick] (1,1)--(1,-2)--(-2,1)--cycle;
\node at (0.5,-0.5) {\Large $A$};
\draw[dashed,thick] (8,1)--(2,1)--(1,2)--(1,8);
\node at (5.5,5.5) {\Large $B$};
\draw[dashed,thick] (0,8)--(0,2)--(-2,2)--(-8,8);
\node at (-2.5,5.5) {\Large $C$};
\draw[dashed,thick] (-8,1)--(-3,1)--(-3,2)--(-8,7);
\node at (-6.5,3.5) {\Large $D$};
\draw[dashed,thick] (0,-8)--(0,-2)--(-2,0)--(-8,0);
\node at (-5.5,-5.5) {\Large $E$};
\draw[dashed,thick] (1,-8)--(1,-3)--(2,-3)--(7,-8);
\node at (3.5,-6.5) {\Large $F$};
\draw[dashed,thick] (8,0)--(2,0)--(2,-2)--(8,-8);
\node at (5.5,-2.5) {\Large $G$};
\draw[-stealth,thick] (-8.5,-2)--(8.5,-2) node [right]{$k$};
\draw[-stealth,thick] (-2,-8.5)--(-2,8.5) node [above]{$l$};
\foreach \n in {-8,...,8}{
\foreach \m in {-8,...,8}{
\draw[fill=black] (\n,\m) circle [radius=\rrad];}}
\draw[fill=blue] (-2,-2) circle [radius=\rad];
\draw[fill=blue] (-2,4) circle [radius=\rad];
\draw[fill=blue] (4,-2) circle [radius=\rad];
\end{tikzpicture}
\caption{Division of poles (green regions), zeroes (blue points) and regular points (black points) of the full partition function into six different regions. After taking into account all cancellations, the residue sums receives contribution from the interior of region A with a -2 factor and from the border of region A with a -1 factor.}
\label{fig.regions}
\end{figure}

To understand how cancellations occur we first divide the poles and zeroes as in \autoref{fig.regions}.
Each different region consists of triples satisfying different relations\footnote{These relations are related to the equivariant version of stability conditions for gauge bundles on $\mathbb{CP}^2$ \cite{Bonelli:2020xps,Klyachko_1990}. Stability conditions have been shown to arise from the projection condition in \autoref{sec.3}. It would be interesting to study if a similar interpretation can be given to the equivariant version.}:
\begin{equation}\begin{split}
    A:\quad& k+l\geq p,\quad k+p\geq l,\quad l+p\geq k,\\
    B:\quad& k+l\geq p,\quad k+p\leq l,\quad l+p\leq k,\\
    C:\quad& k+l\geq p,\quad k+p\geq l,\quad l+p\leq k,\\
    D:\quad& k+l\leq p,\quad k+p\geq l,\quad l+p\leq k,\\
    E:\quad& k+l\leq p,\quad k+p\geq l,\quad l+p\geq k,\\
    F:\quad& k+l\leq p,\quad k+p\leq l,\quad l+p\geq k,\\
    G:\quad& k+l\geq p,\quad k+p\leq l,\quad l+p\geq k.
\end{split}\end{equation}
Note that points along the dashed line saturate one of the inequalities. Recalling that we are restricting to $2\mathfrak{m}_1=k+l+p>0$, these relations determine what are the $k,l,p$ that we can flip without changing the sign of $\mathfrak{m}_1$:
\begin{equation}\begin{split}
    A:\quad& k,l,p,\\
    B:\quad& p,(k,p),(l,p),\\
    C:\quad& l,p,(l,p),\\
    D:\quad& k,(k,l),(k,p),\\
    E:\quad& k,l,(k,l),\\
    F:\quad& l,(k,l),(l,p),\\
    G:\quad& k,p,(k,p).
\end{split}\end{equation}
The brackets, for instance $(l,p)$, mean that both entries can be flipped while keeping $\mathfrak{m}_1$ positive. For reasons that will be explained below, triples in the interior $\mathring{A}$ and in $A-\mathring{A}$ are denoted, respectively, stable and semi-stable contributions. All other points are unstable contributions.

Crucially, flipping the sign of, respectively, $p,l,k$ for a pole in the regions $B$, $D$, $F$ gives a contribution of a pole in $\mathring{A}$  at a larger value of $\mathfrak{m}_1$. Conversely, flipping a sign of a pole in $\mathring{A}$ gives three contributions, one pole for each region $B$, $D$, $F$, at a smaller value of $\mathfrak{m}_1$. We denote $\boldsymbol{k}_{\mathring{A}}$, the orbit given by all triples $(k,l,p)$ which can be reached by a triple in the region $\mathring{A}$. Hence, using \eqref{eq.abstruse}, we find 
\begin{equation}\label{eq.stableRes}
    \sum_{(k,l,p)\in \boldsymbol{k}_{\mathring{A}}}\text{Res}_{a=0}\,Z^\text{full}_{\mathbb{CP}^2}\big|_{(k,l,p)}=-2\text{Res}_{a=0}\,Z_{\mathbb{CP}^2}^\text{full}\big|_{(k,l,p)\in\mathring{A}}.
\end{equation}
Poles in $A-\mathring{A}$, that is the dashed line of the region $A$, are such that the sign of $k,l,p$ can only be flipped twice without changing the sign of\footnote{For the remaining contribution one would find $\mathfrak{m}_1=0$ which is excluded from the sum.} $\mathfrak{m}_1$. Denoting $\boldsymbol{k}_{A-\mathring{A}}$ the corresponding orbit, we find 
\begin{equation}\label{eq.semistableRes}
    \sum_{(k,l,p)\in \boldsymbol{k}_{A-\mathring{A}}}\text{Res}_{a=0}\,Z_{\mathbb{CP}^2}^\text{full}\big|_{(k,l,p)}=-\text{Res}_{a=0}\,Z_{\mathbb{CP}^2}^\text{full}\big|_{(k,l,p)\in A-\mathring{A}}.
\end{equation}
We can now discuss the remaining regions $C$, $E$, $G$. In these regions flipping the sign of $k,l,p$ at a pole gives 3 other poles appearing at higher values of $\mathfrak{m}_1$, but in the same region. For example, denoting $\boldsymbol{k}_C$ the orbit of a triple $(k,l,p)\in C$, one finds
\begin{equation}\label{eq.unstableRes}
    \sum_{(k,l,p)\in \boldsymbol{k}_C}\text{Res}_{a=0}\,Z_{\mathbb{CP}^2}^\text{full}\big|_{(k,l,p)}=0.
\end{equation}
The same holds for the regions $E$ and $G$. Finally, zeroes cannot flip as two out $k,l,p$ vanish and regular points flip to other regular points.

Hence, summing \eqref{eq.stableRes}-\eqref{eq.semistableRes}-\eqref{eq.unstableRes} one finds that at each flux sector we simply have to consider the contributions from the interior $\mathring{A}$ (i.e. stable points) with a factor of $-2$ and those in $A-\mathring{A}$ (i.e. semi-stable points) with a factor of $-1$. This restriction conforms to the equivariant version of slope stability \cite{Klyachko_1990,Bonelli:2020xps}.

\subsection{Explicit Computations}
We now have all ingredients required to tackle the sum over residues in \eqref{eq.sumresidue} and compute the expectation value of the observable descending from 5d, with a position dependent 4d coupling.

\paragraph{5d Observable.}
Putting together all the contributions, the partition function of our $\mathcal{N}=2$ $SU(2)$ theory on $\mathbb{CP}^2$ is
\begin{equation}\begin{split}\label{eq.finalres}
    \mathcal{Z}[\mathcal{T}_{\mathbb{CP}^2}]=&\sum_{\mathfrak{m}_1>0}\left(-2\sum_{(k,l)\in\mathring{A}}-\sum_{(k,l)\in A-\mathring{A}}\right)q^{\frac{\hat{a}^2}{\omega_1\omega_2\omega_3}}q^{\omega_1k\,l+\omega_2l\,p+\omega_3p\,k}\\
    &\prod_{(n_1,n_2)\in\mathcal{B}_{\mathfrak{m}_1}\atop(n_1,n_2)\neq(k,l)}\prod_{(n_1,n_2)\in\mathcal{B}^\circ_{\mathfrak{m}_1}\atop(n_1,n_2)\neq(k,l)}\left(\left(n_1-k\right)\epsilon_1+\left(n_2-l\right)\epsilon_2\right)\\
    &\prod_{i=-k+1}^{k}\prod_{j=-l+1\atop (i,j)\neq (0,0)}^{l}\frac{1}{i\epsilon_1+j\epsilon_2}\prod_{i=-l+1}^{l}\prod_{j=-p+1\atop (i,j)\neq (0,0)}^{p}\frac{1}{i(\epsilon_2-\epsilon_1)+j(-\epsilon_1)}\\
    &\prod_{i=-p+1}^{p}\prod_{j=-k+1\atop (i,j)\neq (0,0)}^{k}\frac{1}{i(-\epsilon_2)+j(\epsilon_1-\epsilon_2)}\\
    &Z^\text{inst}_{\mathbb{C}^2}\left(k\epsilon_1-l\epsilon_2|q_1,\epsilon_1,\epsilon_2\right)Z^\text{inst}_{\mathbb{C}^2}\left(l(\epsilon_2-\epsilon_1)-p(-\epsilon_1)|q_2,\epsilon_2-\epsilon_1,-\epsilon_1\right)\\
    &Z^\text{inst}_{\mathbb{C}^2}\left(p(-\epsilon_2)-k(\epsilon_1-\epsilon_2)|q_3,-\epsilon_2,\epsilon_1-  \epsilon_2\right).
\end{split}\end{equation}
Here, the summation is over pairs $(k,l)$ which belong to the region $A$, corresponding to (semi-)stable contributions. The factors proportional to $q$ in the first line arise from the classical part \eqref{eq.class.CP2} evaluated on \eqref{eq.poles.US} and from the rewriting of the instanton partition function \eqref{eq.instCP2}. The second line is the one-loop determinant \eqref{eq.1loopCP2} from which one excludes the location of the pole $(n_1,n_2)=(k,l)$. The third and fourth lines arise from $R^{\ell}_{n_\ell n_{\ell+1}}$ in \eqref{eq.instCP2}, where $(n_1,n_2,n_3)=(k,l,p)$ is determined by the value of the pole under consideration. With respect to the definition in \eqref{eq.Rn1n2}, we include the contribution at the denominator in \eqref{eq.instCP2} which does not vanish at the location of the pole. This instructs us to remove only the contribution $(0,0)$ from the product. Finally, also the last two lines arise from \eqref{eq.instCP2} and they give rise to an expansion in powers of $q$.

All poles with $\mathfrak{m}_1=1$ are unstable and thus do not contribute. The first non-trivial terms arise at $\mathfrak{m}_1=2$. These are semi-stable points coming with a factor of -1
\begin{equation}
    (k,l,p)=(1,1,2),(2,1,1),(1,2,1),
\end{equation}
and contributing as follows to the partition function
\begin{equation}\begin{split}\label{eq.qexp}
    Z_{\mathbb{CP}^2}=&-q^{-\frac{\left(2-\frac{1}{2}(\omega_1-1)\right)^2}{\omega_1\omega_2\omega_3}}\frac{q^{\omega_1}q^{2\omega_2}q^{2\omega_3}(\epsilon_1+\epsilon_2)}{2\epsilon_1\epsilon_2}+q^{-\frac{\left(2+\frac{1}{2}(\omega_2-1)\right)^2}{\omega_1\omega_2\omega_3}}\frac{q^{2\omega_1}q^{\omega_2}q^{2\omega_3}(2\epsilon_1-\epsilon_2)}{2\epsilon_1(\epsilon_1-\epsilon_2)}\\
    &+q^{-\frac{\left(2+\frac{1}{2}(\omega_3-1)\right)^2}{\omega_1\omega_2\omega_3}}\frac{q^{2\omega_1}q^{2\omega_2}q^{\omega_3}(\epsilon_1-2\epsilon_2)}{2\epsilon_2(\epsilon_1-\epsilon_2)}+\dots.
\end{split}\end{equation}
Terms with $\mathfrak{m}_1>2$ contribute at a higher power of $q$.

As discussed before, these equivariant invariants do not match those appearing in \cite{Bershtein:2015xfa,Bonelli:2020xps} and they have not, to our knowledge, appeared previously in the literature. However, there is nothing which prevents us from turning on  the observable in \cite{Bershtein:2015xfa,Bonelli:2020xps}, and turning off the one coming from 5d. In this case we would reproduce $SU(2)$ equivariant Donaldson invariants \cite{Gottsche:2006tn,Bershtein:2015xfa,Bonelli:2020xps}.

\paragraph{Non-Equivariant Limit.}
As we argued above, in the non-equivariant limit\footnote{Recall that, in this limit, on top of $\epsilon_1,\epsilon_2\rightarrow 0$, we also take $\omega_i\rightarrow 1$.} also the classical contribution reproduces that in \cite{Bershtein:2015xfa,Bonelli:2020xps}. However, in order to reproduce Donaldson invariants we also have to turn on a specific observable and take the non-equivariant limit. The observable is \cite{Bershtein:2015xfa,Bonelli:2020xps}
\begin{equation}
    \exp\left(\sum_{\ell=1}^3(2a+2\omega_\ell\mathfrak{m}_1)^2\Omega_\ell\right),
\end{equation}
where
\begin{equation}
    \Omega_\ell=(0,-z\epsilon_1+x(\epsilon_1)^2,-z\epsilon_2+x(\epsilon_2)^2).
\end{equation}
Considering the first terms in \eqref{eq.qexp}, one finds the following expansion in $q$
\begin{equation}\label{eq.qexp.nonequiv}
    Z_{\mathbb{CP}^2}^\text{non-equiv}=-\frac{3}{2}qz+\mathcal{O}(q^2).
\end{equation}
All other terms in \eqref{eq.qexp} contribute at a higher power in $q$. The result in \eqref{eq.qexp.nonequiv} reproduces expressions for Donaldson invariants in the literature \cite{ellingsrud1995wallcrossing}.

\section{Discussion}\label{sec.5}
In this work, we have performed the integral over the Coulomb branch parameter of the partition function of an $\mathcal{N}=2$ topologically twisted theory on $\mathbb{CP}^2$ with $SU(2)$ gauge group. Our expression, which descends from the partition function of an $\mathcal{N}=1$ theory on $S^5$, is in terms of a single physical flux, rather that three equivariant fluxes. This is compensated by a larger sum over residues at each topological sector. Taking this into account, the structure of the residue sum precisely reproduces that in \cite{Bershtein:2015xfa,Bershtein:2016mxz,Bonelli:2020xps}. The only difference arises as the 4d coupling constant obtained dimensionally reducing from $S^5$ is position dependent. This defines a new observable and, in this work, we have computed its expectation value\footnote{Let us stress once more that, as both observable are supersymmetric, we could in principle turn off the one from 5d and turn on the one with a constant coupling constant. Doing so we would precisely reproduce the results in \cite{Bershtein:2015xfa,Bershtein:2016mxz,Bonelli:2020xps}.}. We have also shown that in the non-equivariant limit our observable reproduces $SU(2)$ Donaldson invariants. While the $SO(3)$ case is not considered in this work, it can be addressed with simple modifications of the computations for the $SU(2)$ case. While for $SU(2)$ the projection condition \eqref{eq.projection.CP2} restrict $t$ to be even, the equivariant invariants computed by the $SO(3)$ theory are obtained by considering only odd values of $t$. In this case only stable contributions arise, weighted by a factor of -2.

\subsection{Future Directions}
This work is just the first stepping stone of a program which aims at computing exactly partition function of generic $\mathcal{N}=2$ theories and to study its physical and mathematical properties. Given the (relative) simplicity of our procedure, achieving the following goals now seems within reach.

\paragraph{Higher rank gauge theories.} The methodology described in this paper can be extended to $SU(N)$ theories. This applies to both the stability condition restricting the sum over fluxes and Zamolodchikov's rewriting of the Nekrasov partition function \cite{Sysoeva:2022syp}. The large $N$ limit of these theories has potential applications to AdS/CFT, see for instance \cite{Hosseini:2018uzp}.

\paragraph{Quasi-toric four-manifolds.} Exploiting the results in \cite{Lundin:2023tzw,Ruggeri:2025kmk} one can compute the integrand of the partition function for topologically twisted theories for a large class of toric four-manifolds arising as $S^1$-quotients of regular Sasakian manifolds. 

For instance, one can study on the four-dimensional manifold arising as a quotient $S^2\times S^2=T^{1,1}/S^1$. In this case, the sum is over two independent physical fluxes or over four equivariant fluxes $(k^1,k^2,k^3,k^4)$ associated to the four divisors of $S^2\times S^2$. A version of the abstruse duality \eqref{eq.abstruse2} has been shown to continue to hold for generic compact toric manifolds \cite{Bonelli:2020xps}. On $S^2\times S^2$ the corresponding expression is
\begin{equation}\begin{split}\label{eq.abstruse3}
    &\text{Res}_{a=0}\,Z_\text{equiv}^\text{full}\big|_{(k^1,k^2,k^3,k^4)}+\text{Res}_{a=0}\,Z_\text{equiv}^\text{full}\big|_{(-k^1,k^2,k^3,k^4)}\\
    &+\text{Res}_{a=0}\,Z_\text{equiv}^\text{full}\big|_{(k^1,-k^2,k^3,k^4)}+\text{Res}_{a=0}\,Z_\text{equiv}^\text{full}\big|_{(-k^1,-k^2,k^3,k^4)}=0,
\end{split}\end{equation}
A difference arising when increasing the number of fixed points $\chi$ is that the poles of the partition function are of order $\chi-2$ \cite{Bonelli:2020xps}. Hence, it is expected that the partition function on $S^2\times S^2$ has double poles. This would affect the computation of the observable \eqref{eq.finalres}.

\paragraph{Orbifolds.} Recently, SQFTs on spaces with orbifold singularities have drawn a lot of attention in connection to supergravity solutions with singularities in their near-horizon geometry \cite{Ferrero:2020twa}. The simplest case to address is a topologically twisted theories on a 4d weighted projective space \cite{Martelli:2023oqk,Mauch:2024uyt,Mauch:2025irx}. This could define new equivariant invariants for four-dimensional orbifolds.

\paragraph{Matter.} Along the lines of \cite{Festuccia:2020yff}, it would be interesting to study the inclusion of matter. In case the manifold (orbifold) is not spin, the introduction of a spin$^c$ connection is required. This would enable us to study $\mathcal{N}=2^*$ theories as in \cite{Pestun:2007rz}.

\paragraph{Exotic theories.} The most exciting avenue of research is to obtain exact results for non-topological theories, which extend Pestun theory on $S^4$ to more generic four-manifolds \cite{Festuccia:2018rew,Festuccia:2019akm}. As mentioned in \autoref{sec.2}, for exotic theories the field strength localizes to instantons $(+)$ and anti-instantons ($-$) at different fixed points. 

The first example one can address is the $SU(2)$ theory on $\mathbb{CP}^2$ \cite{Lundin:2021zeb} with a $-++$ distribution of instantons and anti-instantons at the three fixed points. By using the relation between the instanton and anti-instanton partition function \cite{Festuccia:2018rew,Mauch:2021fgc}
\begin{equation}
    \mathcal{Z}_{\mathbb{C}^2}^\text{anti-inst}(a|\bar{q},\epsilon_1,\epsilon_2)=\mathcal{Z}_{\mathbb{C}^2}^\text{inst}(a|\bar{q},-\epsilon_1,\epsilon_2)=\mathcal{Z}_{\mathbb{C}^2}^\text{inst}(a|\bar{q},\epsilon_1,-\epsilon_2)
\end{equation}
one can rewrite the anti-instantons partition function as one for instantons with a flipped equivariant parameter. Effectively one then obtains a theory on $\mathbb{CP}^2$ where the local equivariant parameter at the first fixed point is given by $-\epsilon_1$.

This affects the pole structure of the full partition function. The partition function $\mathcal{Z}_{\mathbb{C}^2}^\text{inst}(a|q,\epsilon_1^\ell,\epsilon_2^\ell)$ has a pole if $\epsilon_1^\ell\epsilon_2^\ell<0$ and is regular otherwise \cite{Bonelli:2020xps}. On $\mathbb{CP}^2$ there is only one patch with $\epsilon_1^\ell\epsilon_2^\ell<0$ and this explains why the partition function for equivariant DW theory only has simple poles. For the exotic theory, there are instead two patches with $\epsilon_1^\ell\epsilon_2^\ell<0$ and thus the partition function will have double poles.

Finally, the abstruse dualities \eqref{eq.abstruse} and \eqref{eq.abstruse3} rely on the fact that $\epsilon_1^\ell\epsilon_2^{\ell+1}<0$ for any compact toric manifold. This will not be the case for the exotic theory where
\begin{equation}
    \epsilon_1^1\epsilon_2^2=(-\epsilon_1)(-\epsilon_1)>0.
\end{equation}
What one expects then is that the abstruse duality \eqref{eq.abstruse} continues to hold when flipping $k^2$ or $k^3$. Instead when one flips $k^1$ the following holds
\begin{equation}
    \text{Res}_{a=0}\,Z_\text{equiv}^\text{full}\big|_{(k^1,k^2,k^3)}=\text{Res}_{a=0}\,Z_\text{equiv}^\text{full}\big|_{(-k^1,k^2,k^3)}.
\end{equation}
By using these relation one can still simplify the sum over residues arising from the poles distribution of the full partition function.

\paragraph{Fibering Operators.} The contour integral for $\mathbb{CP}^2$ can be uplifted to $\mathcal{N}=1$ theories on $\mathbb{CP}^2\times S^1$ and $S^5$. In the former case one has to take into consideration the compactification of the contour integral over $a$. On $S^5$, instead, the contour of integration remains the same\footnote{We derived our contour of integration on $\mathbb{CP}^2$ from that on $S^5$ \cite{Qiu:2016dyj}.} and poles at different values of $\mathfrak{m}$ will contribute at the same topological sector but at different values of $t$. Moreover, exploiting the fibering operator in \cite{Closset:2018ghr,Closset:2022vjj}, the partition function on $S^5$ (and more in general on $S^5/\mathbb{Z}_h$) should be computable starting from that on $\mathbb{CP}^2\times S^1$.

\paragraph*{Acknowledgments}
We would like to thank Roman Mauch and Nicolò Piazzalunga for useful discussions and initial collaboration on this project. We also thank Guido Festuccia, Massimiliano Ronzani and Itamar Yaakov for stimulating discussions on the subject. LR acknowledges partial support by the INFN.

\bibliographystyle{utphys}
\bibliography{main}

\end{document}